\documentclass[traditabstract]{aa}

\usepackage{graphicx}
\usepackage{txfonts}
\usepackage{natbib}
\usepackage[colorlinks=true,citecolor=blue]{hyperref}
\usepackage{xspace}
\usepackage{longtable}
\usepackage{array}
\usepackage{lscape}
\usepackage{rotating}
\usepackage[labelformat=simple]{caption,subcaption}
\usepackage{ifthen}
\usepackage{amsmath,amssymb}
\usepackage{bm}% to write bold letters in math mode     

\bibpunct{(}{)}{;}{a}{}{,}

\newcommand{\mic}{$\mu$m\xspace}
\newcommand{\as}{\hbox{$^{\prime\prime}$}\xspace}
\newcommand{\lsd}{\hbox{$\lambda/D$}\xspace}
\newcommand{\cpup}{c/p\xspace}

\defcitealias{N'Diaye2013}{Paper I}
\defcitealias{N'Diaye2016}{Paper II}
\usepackage[normalem]{ulem}

\begin{document}

\title{Calibration of quasi-static aberrations in exoplanet direct-imaging instruments with a Zernike phase-mask sensor\\ III. On-sky validation in VLT/SPHERE}
\titlerunning{Zernike phase-mask sensor. III.}

\author{
    A. Vigan\inst{1} 
    \and
    M. N'Diaye\inst{2}
    \and
    K. Dohlen\inst{1}
    \and
    J.-F. Sauvage\inst{1,3}
    \and
    J. Milli\inst{4}
    \and
    G. Zins\inst{4}
    \and
    C. Petit\inst{3}
    \and\\ 
    Z. Wahhaj\inst{4}
    \and
    F. Cantalloube\inst{5}
    \and
    A. Caillat\inst{1}
    \and
    A. Costille\inst{1}
    \and
    J. Le Merrer\inst{1}
    \and
    A. Carlotti\inst{6}
    \and
    J.-L. Beuzit\inst{1}
    \and
    D. Mouillet\inst{6}
}

\institute{
    Aix Marseille Univ, CNRS, CNES, LAM, Marseille, France %1
    \\ \email{\href{mailto:arthur.vigan@lam.fr}{arthur.vigan@lam.fr}}
    \and
    Universit\'e C\^ote d'Azur, Observatoire de la C\^ote d'Azur, CNRS, Laboratoire Lagrange, France %2
    \and
    ONERA, The French Aerospace Lab, BP72, 29 avenue de la Division Leclerc, 92322 Ch\^{a}tillon Cedex, France %3
    \and
    European Southern Observatory, Alonso de Cordova 3107, Vitacura, Santiago, Chile %4
    \and 
    Max Planck Institute for Astronomy, K\"onigstuhl 17, 69117, Heidelberg, Germany %5
    \and
    Univ. Grenoble Alpes, CNRS, IPAG, F-38000 Grenoble, France %6
}

\date{Received 15 May 2019 / Accepted 18 July 2019}

\abstract{
  Second-generation exoplanet imagers using extreme adaptive optics (ExAO) and coronagraphy have demonstrated their great potential for studying close circumstellar environments and for detecting new companions and helping to understand their physical properties. However, at very small angular separation, their performance in contrast is limited by several factors: diffraction by the complex telescope pupil (central obscuration and spiders) not perfectly canceled by the coronagraph, residual dynamic wavefront errors, chromatic wavefront errors, and wavefront errors resulting from noncommon path aberrations (NCPAs). These latter are differential aberrations between the visible wavefront sensing path of the ExAO system and the near-infrared science path in which the coronagraph is located. In a previous work, we demonstrated the use of a Zernike wavefront sensor called ZELDA for sensing NCPAs in the VLT/SPHERE exoplanet imager and their compensation with the high-order deformable mirror of the instrument. These early tests on the internal light source led to encouraging results for the attenuation of the quasi-static speckles at very small separation. In the present work, we move to the next step with the on-sky validation of NCPA compensation with ZELDA. With an improved procedure for the compensation of NCPAs, we start by reproducing previous results on the internal source. We show that the amount of aberration integrated between 1 and 15~cycles/pupil (\cpup) is decreased by a factor of approximately five, which translates into a gain in raw contrast of between 2 and 3 at separations below 300 mas. On sky, we demonstrate that NCPA compensation works in closed loop, leading to an attenuation of the amount of aberration by a factor of approximately two. However, we identify a loss of sensitivity for the sensor that is only partly explained by the difference in Strehl ratio between the internal and on-sky measurements. Our simulations show that the impact of ExAO residuals on ZELDA measurements is negligible for integration times beyond a few tenths of a second. Coronagraphic imaging on sky is improved in raw contrast by a factor of 2.5 at most in the ExAO-corrected region. We use coronagraphic image reconstruction based on a detailed model of the instrument to demonstrate that both internal and on-sky raw contrasts can be precisely explained, and we establish that the observed performance after NCPA compensation is no longer limited by an improper compensation for aberration but by the current apodized-pupil Lyot coronagraph design. We finally conclude that a coronagraph upgrade combined to a proper NCPA compensation scheme could easily bring a gain in raw contrast of a factor of two to three below 200~mas.}


\maketitle

\section{Introduction}
\label{sec:introduction}

With the advent of the second-generation, high-contrast instruments on the ground in 2013-2014, unprecedented performance has been obtained on the images of nearby stars with contrasts down to $10^{-6}$ at separations beyond 300\,mas in the near-infrared (NIR) band, leading to the observation of various circumstellar disks \citep[e.g.,][]{deBoer2016,Lagrange2016,Currie2017,Feldt2017,Goebel2018,Esposito2018} and the discovery of new young gas giant planets \citep[e.g.,][]{Macintosh2015,Chauvin2017,Keppler2018}. These results begin to shed light on the architecture of planetary systems, planet formation and evolution, and atmospheric properties of young giant planets and brown dwarfs. In terms of planet demography, NIR surveys of hundreds of nearby stars have so far shown that giant gaseous planets on orbits wider than 10 AU remain rare, mostly due to inefficient formation of such companions at large distances \citep[e.g.,][]{Bowler2016,Uyama2017,Vigan2017,Chauvin2017b,Nielsen2019arXiv}. 

Imaging planets on shorter orbits from the ground requires observations at closer separations and deeper contrasts than what is currently done. A first step towards this challenging objective consists in thoroughly understanding the performance and limitations of the latest exoplanet imagers, such as Gemini/GPI, Subaru/SCExAO, and VLT/SPHERE (\citealt{Macintosh2014}; ~ ~ ~ ~ ~ \citealt{Jovanovic2015}; \citealt{Beuzit2019}). These advanced instruments use a common recipe based on extreme adaptive optics (ExAO) to correct for the effects of the atmospheric turbulence on the incoming wavefront, coronagraphy to remove starlight due to the telescope diffraction effects on the star image, and post-processing methods to reduce the residual scattered light in the coronagraphic image and detect the signal of a companion or a disk.

Even in good observing conditions, the $10^{-6}$ contrast in coronagraphic images can be quickly degraded by many effects, such as residual jitter of a few milliarseconds (mas), extremely low-order aberrations (tip, tilt and defocus), the low-wind effect \citep{Sauvage2016}, or the wind-driven halo \citep{Cantalloube2018}. Over the past few years, various strategies have been investigated to mitigate these issues \citep[e.g.,][]{Baudoz2010,Singh2015,Huby2015,Huby2017,Lamb2017,Milli2018,N'Diaye2018,Wilby2018}. In the absence of or after compensation for these effects, there are still other hurdles, the most infamous being ``speckles'', which are residual scattered light in the coronagraphic images that prevent the observation of the close stellar environment. These artifacts first come from the wavefront errors due to the atmospheric turbulence residuals after ExAO correction. Different upgrades are under investigation to push the limits of the AO performance further, such as new or additional deformable mirrors, high-sensitivity wavefront sensors, control loops with frequency faster than 1\,kHz, and predictive control algorithms \citep[e.g.,][]{Fusco2016,Chilcote2018,Guyon2018}. 

Other speckles originate from the wavefront errors that are invisible to the ExAO system, the so-called noncommon path aberrations (NCPAs). Anticipated at the time of design of the exoplanet imagers \citep[e.g.,][]{Fusco2006}, these wavefront errors are due to the differential optical path between the ExAO visible wavefront sensing and the NIR science camera arms. These aberrations slowly evolve with time, producing speckles with a quasi-static behavior that makes their calibration challenging for exoplanet imaging \citep{Martinez2012,Martinez2013,Milli2016}. In addition to post-processing techniques, this issue is currently being addressed thanks to the exploration of several solutions including online software \citep[e.g.,][]{Savransky2012,Martinache2014,Bottom2017} and hardware \citep[e.g.,][]{Galicher2010,Paul2013,Paul2014,Martinache2013,Martinache2016,Wilby2017}. Among these solutions, we proposed the use of a Zernike wavefront sensor to calibrate NCPAs and enable deeper contrast at closer angular separation.

The ZELDA sensor is based on phase-contrast techniques, originally proposed by \citet{Zernike1934}, to measure NCPAs in high-contrast imaging instruments with nanometric accuracy. This sensor uses a focal plane phase mask to produce interference between a reference wave created by the mask and the phase errors present in the system. As a result, this sensor converts the aberrations in the entrance pupil into intensity variations in the exit pupil. This phase-to-intensity conversion depends on the mask characteristics, that is, the diameter and the depth that is related to the sensor phase delay. In \citet{N'Diaye2013}, hereafter \citetalias{N'Diaye2013}, we established the formalism for this approach, showing its ability to measure static aberration with sub-nanometric accuracy and deriving theoretical contrast gains after compensation for exoplanet observation. This approach has also been studied in the context of phase discontinuities such as segmented apertures \citep{Janin-Potiron2017} or low-wind effects \citep{Sauvage2016spie}.

In \citet{N'Diaye2016}, hereafter \citetalias{N'Diaye2016}, we reported the validation of our method with the implementation of a prototype called ZELDA (Zernike sensor for Extremely Low-level Differential Aberration) on VLT/SPHERE. The first results on the internal source demonstrated a significant contrast gain in the coronagraphic image after compensation for NCPAs using the deformable mirror of the 
instrument. Following our successful experiment, we now move to the next step with on-sky NCPA measurement and compensation.

Noncommon path aberration compensation can be addressed in two different ways: (i) full compensation on sky or (ii) NCPA calibration on the internal source and application of the correction on sky. Although the second strategy is obviously more efficient and avoids dealing with problems such as integration times and variable observing conditions, it also requires a complete understanding of the behavior of the instrument when switching between the internal calibration source and a star. It also relies on the presence of a calibration source sufficiently close to the entrance of the instrument, which is the case in VLT/SPHERE, but even so the light will not see the telescope mirrors as there is rarely (if ever) an appropriate light source in the telescope itself. Although theoretically applicable, this strategy was deemed too risky for our first on-sky tests and we adopted the first approach instead.

To carry out this demonstration, ESO awarded us three technical half-nights on 2018-04-01, 2018-04-02, and 2018-04-03\footnote{Throughout this work we use the ISO date format, i.e. the YYYY-MM-DD notation}. The observing conditions were unfortunately quite unstable on April 1 and 3, and very poor on April 2. In this paper, we report on the on-sky tests of NCPA calibration with ZELDA on VLT/SPHERE with the acquired data on the first and third half-nights. We first detail our implemented improvements in the wavefront error calibration and compensation with our Zernike sensor in the instrument compared with \citetalias{N'Diaye2016}. The on-sky results are then presented with a comprehensive analysis to validate our approach. We finally quantify the impact of NCPA correction on coronagraphic data and discuss prospects for further exoplanet direct imaging and spectroscopy observations. 

%--------------------------------------------------------------------------------------------------------------------
\section{Improvements in the NCPA calibration}
\label{sec:improvement_ncpa_calibration}

The first attempt presented in \citetalias{N'Diaye2016} was a working demonstration but had a few shortcomings that made it difficult to implement in a robust and repeatable way. 

The first one was the low-pass filtering of the spatial frequencies contained in the ZELDA optical path difference (OPD) maps. This filtering is a mandatory step because ZELDA measures wavefront errors with spatial frequencies up to 192~\cpup (the pupil image has 384 pixels in diameter on the detector), while the SPHERE high-order deformable mirror (HODM) only corrects for the spatial frequencies up to 20~\cpup. In \citetalias{N'Diaye2016}, we implemented a simple Hann filtering in the Fourier space with a radius of 25~\cpup. This approach proved very efficient at removing the high-spatial frequencies of the OPD maps, but the filter radius was arbitrary and disconnected from the 990 Karhunen-Lo\`eve (KL) modes that are used by SAXO, the SPHERE ExAO system, in its control architecture \citep{Petit2008a}. As a result, the filtered OPD maps may still contain spatial frequencies that cannot be controlled, and therefore cannot be corrected for by the system.

To improve the spatial filtering we implemented a new approach based on the control matrices of SAXO. The unfiltered OPD maps are first directly converted into HODM actuator voltages before being projected onto the KL modes of the system. At this level, this projection has already removed all the spatial frequencies that cannot be seen and controlled by SAXO. On top of that, a user-defined number of additional modes can be removed by setting their value to zero in the projection; this is done to restrict the NCPA correction only to the lowest spatial frequencies. These filtered modes are then converted back to voltages and finally converted into HOWFS reference slope offsets. This procedure has the powerful advantage that it takes into account all the specificity of the system, and in particular defects like dead actuators, the knowledge of which is incorporated into the control architecture of SAXO.

The second shortcoming was the centering of the source point spread function (PSF) on the ZELDA mask. We first remind that the tip-tilt control in SPHERE is handled in two stages \citep{Sauvage2016}: first with a fast (800~Hz bandwidth) tip-tilt mirror that provides an rms residual jitter of  2~mas in the visible in nominal conditions, and then with the slow (1~Hz) differential tip-tilt sensor (DTTS) that is implemented in the NIR arm to compensate for the differential chromatic tip-tilt between the visible and the NIR \citep{Baudoz2010}. The goal of the DTTS loop is to provide a fine (<0.5~mas) stabilization of the PSF on the coronagraph mask, or in this case on the ZELDA mask. The centering of the PSF on our sensor spot was previously handled by hand, that is, by manually changing the reference slopes of the DTTS until the residual tip-tilt in the ZELDA signal, estimated by eye only, is deemed negligible. To improve this situation, the tip-tilt in now estimated from the OPD maps by projecting them on the tip (Z1) and tilt (Z2) Zernike polynomials computed for the SPHERE pupil (taking into account the 14\% central obscuration and the HODM dead actuators). The tip-tilt is then converted into an offset on the DTTS reference slopes, which are updated accordingly to center the PSF on the mask. Before proceeding to the correction for the higher-order NCPA as described previously, the tip and tilt terms are subtracted from the OPD maps.

Finally, we also implemented several improvements to the ZELDA analysis code. Based on the IDL code that was originally developed for \citetalias{N'Diaye2013}, we developed a new Python-based generic code called \texttt{\href{https://github.com/avigan/pyZELDA}{pyZELDA}} that is dedicated to the analysis of data acquired with Zernike WFS \citep{Vigan2018ascl}. This code is public and freely available under the MIT license\footnote{\url{https://github.com/avigan/pyZELDA}}. It is easily extensible to include new instruments or laboratory test beds.

All these improvements make NCPA calibration and compensation with ZELDA much more robust. As a result, NCPA calibration has been transposed into a calibration template that is executed in the SPHERE daily calibration plan to monitor the long-term evolution of NCPAs. The stability of the aberrations in SPHERE has begun to be explored in the laboratory \citep{Martinez2013} and in Paranal \citep{Milli2016} based on focal-plane coronagraphic images, but these measurements do not constitute a direct measurement of the phase errors. The short- and long-term stability of NCPAs that are based on the ZELDA phase measurements will be explored in a future study.

%--------------------------------------------------------------------------------------------------------------------
\section{On-sky calibration and compensation}
\label{sec:on_sky_ncpa_calibration}

There are two complementary approaches to analyze the NCPA compensation in high-contrast imaging: the first one is to look at the wavefront measurements provided by the NCPA WFS (be it ZELDA or a different one), and the second is to estimate the raw contrast gain in coronagraphic images. While the most important parameter in the end is the final gain in raw contrast, the first approach is essential to understanding the instrument. In this section we use this first approach to investigate the behavior and limitations of ZELDA in the presence of ExAO-filtered atmospheric residuals.

\subsection{Description of the tests}
\label{sec:decription_tests_ncpa}

\begin{table*}
    \caption[]{NCPA measurement log}
    \label{tab:ncpa_loop_log}
    \centering
    \begin{tabular}{lccccccc}
    \hline\hline
    Test name & Date         & UTC time  & Source       & DIMM seeing     & $\tau_{0}$    & $N_{\mathrm{iter}}$ & DIT$\times$NDIT\tablefootmark{a} \\
              &              &           &              & (\as)           & (ms)          &                     & (s$\times$\#) \\ 
    \hline
    % NCPA compensation
    ZT01      & 2018-04-01   & 19:05     & Internal     &                 &               & 4                   & 1$\times$10   \\
    ZT02      & 2018-04-02   & 02:41     & $\alpha$ Crt & 0.71 $\pm$ 0.07 & 5.0 $\pm$ 0.4 & 4                   & 7$\times$4    \\
    \hline
    % NCPA compensation
    ZT03      & 2018-04-03   & 17:17     & Internal     &                 &               & 4                   & 1$\times$10   \\
    ZT04      & 2018-04-04   & 00:23     & 1 Pup        & 0.77 $\pm$ 0.05 & 4.0 $\pm$ 0.4 & 3                   & 3$\times$10   \\
    % sky vs. internal comparison
    ZT05      & 2018-04-04   & 01:04     & 1 Pup        & $\sim$0.80      & $\sim$3.0     & 0                   & 3$\times$10   \\
    ZT06      & 2018-04-04   & 01:33     & Internal     &                 &               & 0                   & 1$\times$10   \\
    \hline
    \end{tabular} 
    \tablefoot{Missing value indicates no applicable value for the test. \tablefoottext{a}{Value for each iteration.}}
\end{table*}

Various tests were performed with ZELDA, both on the internal point source of SPHERE located in the instrument calibration unit \citep{Wildi2009,Wildi2010}, and on bright single stars that were selected in advance to ensure optimal ExAO performance \citep{Sauvage2016}. Indeed, the primary goal of these tests was to validate ZELDA in optimal conditions rather than explore the sensor flux limit. The two stars that were used are $\alpha$~Crt ($R=3.27$, $H=1.76$) and 1~Pup ($R=3.29$, $H=0.93$). The different tests that are relevant for our analysis are summarized in Table~\ref{tab:ncpa_loop_log}. Unless otherwise stated, all tests are started with the HOWFS reference slopes calibrated during the SPHERE daily calibration sequence.

The internal source and on-sky tests are handled in mostly the same way, except for the initial setup of the light source. On the internal source, the calibration unit is set up to provide a NIR point source, while on sky the full target acquisition procedure is used to point the star, engage the telescope active optics and guiding, and setup the instrument. In either case, the system ends up with all AO loops closed, providing a diffraction-limited PSF. A specific setup is then manually handled in the instrument: ZELDA mask in the coronagraph wheel, and pupil imaging mode with \texttt{N\_FeII} filter ($\lambda = 1642$~nm, $\Delta\lambda = 24$~nm) in the IRDIS camera (see \citetalias{N'Diaye2016} for more details). At this level, the PSF falls close enough to the center of the ZELDA mask to provide a measurable signal. The detector integration time (DIT) is then adjusted so that the average flux in the pupil is approximately at half the linearity range of the Hawaii-2RG detector.

The ZELDA analysis requires two calibration pupil images \citepalias{N'Diaye2016}. The first one is an instrumental background, which is obtained by opening the AO loops, closing the  entrance shutter of the instrument, acquiring an image with IRDIS, and reverting back to the original instrument state. The second is a clear pupil image obtained by moving the ZELDA mask out of the PSF with a small rotation of the wheel that holds the mask, acquiring an image with IRDIS and again reverting back to the original state. Once these two calibrations have been acquired, the ZELDA mask is no longer moved and the NCPA calibration and compensation can start.

The NCPA compensation is implemented as an iterative process for reasons that become clear in Sect.~\ref{sec:zelda_sensitivity}. At each iteration, a pupil image is acquired with IRDIS and analyzed with \texttt{pyZELDA} to produce an OPD map calibrated in nanometers. In this analysis, the pupil features (central obscuration, spiders) are taken into account to compute the wavefront reconstruction. The obscured parts are, in the end, masked numerically because the ZELDA signal can only be computed in the illuminated parts of the pupil. Moreover, because the reconstruction is performed independently for each individual point in the pupil, the dead actuators of the SPHERE HODM are not specifically taken into account in the reconstruction. However, they produce wavefront errors that are significantly beyond the linear range of the ZELDA sensor, creating a meaningless signal. They are therefore masked numerically in our reconstruction.

Finally, the OPD map is spatially filtered (see Sect.~\ref{sec:improvement_ncpa_calibration}) and the reference slopes of the HOWFS are updated. For the filtering, we conservatively kept 700 modes (out of 990), which corresponds to a low-pass filter with a cutoff at $\sim$15~\cpup. Higher values are possible but the higher modes are noisier and can potentially lead to instabilities. In most cases three or four iterations were performed. More iterations were attempted but beyond four we noticed some instabilities in the AO loop induced by the presence of dead HODM actuators on the right edge of the pupil.

\subsection{Results}
\label{sec:results_ncpa}

\begin{figure*}
    \centering 
    \includegraphics[width=0.49\textwidth]{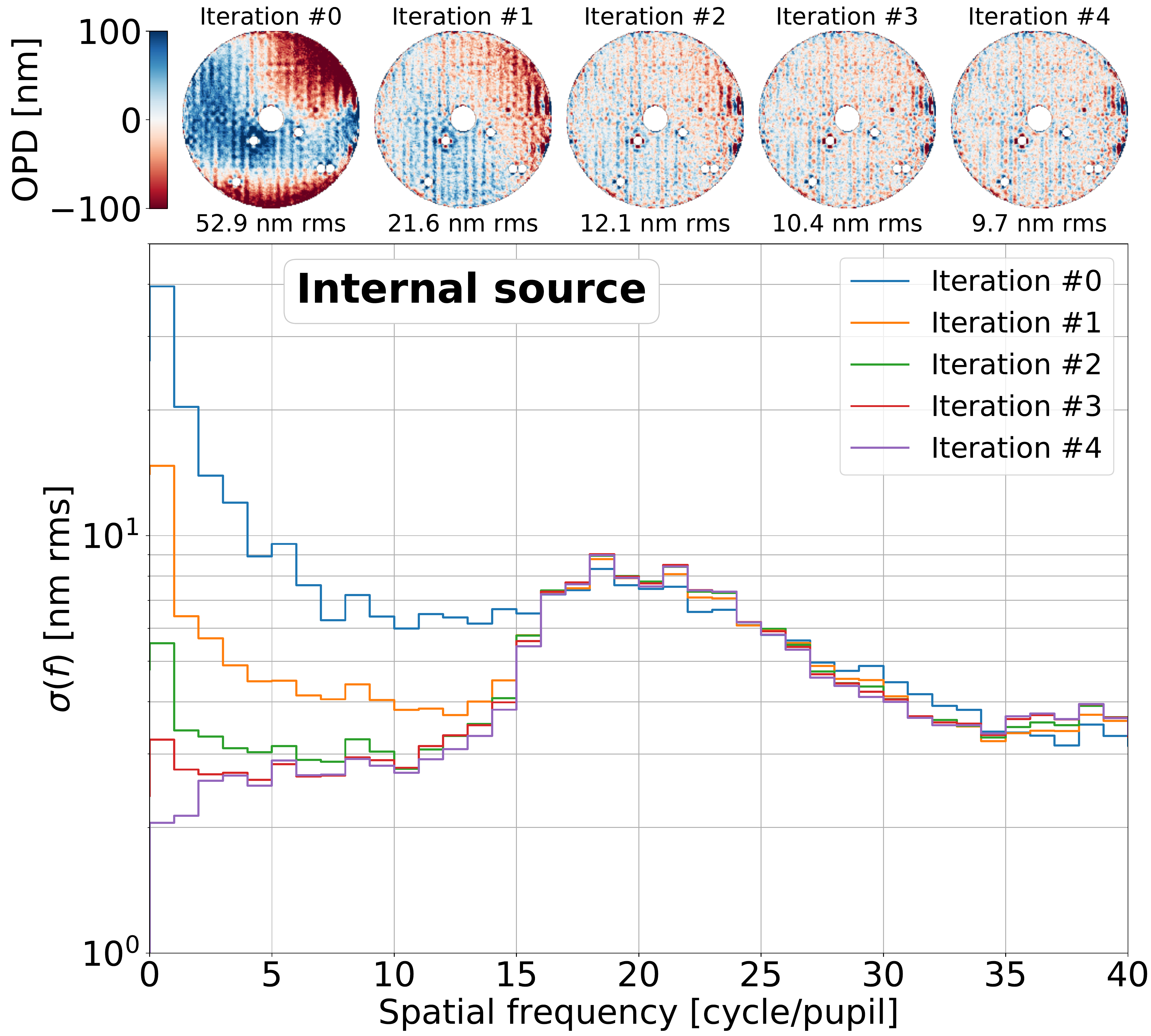}
    \includegraphics[width=0.49\textwidth]{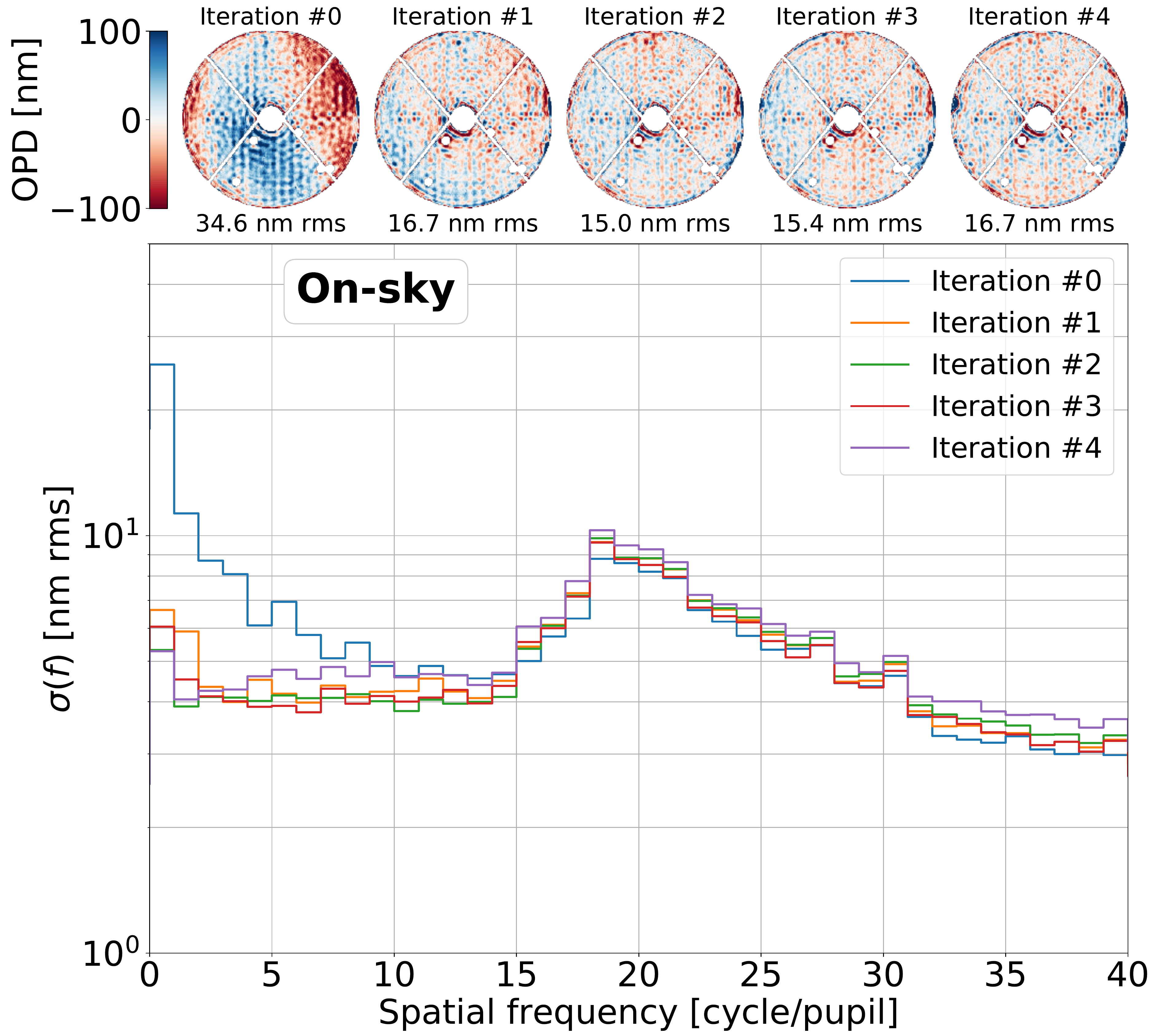}
    \caption{NCPA compensation loop results on 2018-04-01 on the internal source (left, test ZT01) and on sky (right, test ZT02). The top row shows the ZELDA OPD maps at the start (iteration \#0) and at the subsequent iterations. The value reported below each OPD map corresponds to the amount of aberration integrated in the range 1-15~\cpup. The bottom plot shows the $\sigma(f)$ of the OPD map at each iteration (see Sect.~\ref{sec:results_ncpa} for the $\sigma(f)$ definition). In addition to the dead actuators and central part of the pupil that are always masked, the spiders are also masked for the on-sky test.}
    \label{fig:ncpa_loop_1}
\end{figure*}

Our results are presented in Fig.~\ref{fig:ncpa_loop_1} for 2018-04-01. The results for 2018-04-03 are almost identical and are provided in Appendix~\ref{sec:additional_results}. For each NCPA compensation loop we present the OPD maps,
calibrated in nanometers, at each iteration, and the standard deviation $\sigma(f)$ decomposed in terms of spatial frequency $f$. This last term is derived from the classical two-dimensional power spectral density (2D PSD) expressed in (nm~/~(\cpup))$^2$:

\begin{equation}
    \sigma(f) = \sqrt{\int_{f}^{f+1}\int_{0}^{2\pi}\mathrm{PSD}(\nu, \theta)\nu d\nu d\eta},
\end{equation}

\noindent with $\nu$ and $\eta$ being the radial and azimuthal coordinates of the spatial frequencies in the 2D PSD. The $\sigma(f)$ values are discrete and expressed in nm rms. An important property of the PSD is that its integral over all spatial frequencies is equal to the variance of the original OPD map. However, the PSD expressed in (nm~/~(\cpup))$^2$ can sometimes be difficult to physically interpret from the instrumental point of view, where we usually think more in terms of standard deviation of the wavefront errors within a given range of spatial frequencies. The $\sigma(f)$ addresses that issue: each of the points represents the standard deviation of the measured OPD maps in bins of size 1~\cpup. The standard deviation over a wider range of spatial frequencies is obtained by quadratically summing the individual bins within that range.

On the internal source, the starting point is generally dominated by a strong astigmatism and residual tip-tilt due to the fact that the PSF is not precisely centered on the mask at the beginning. The first iteration already brings a major improvement at all spatial frequencies up to $\sim$15~\cpup (because of the mode filtering; see Sect.~\ref{sec:improvement_ncpa_calibration}), but there are still some residual low frequencies, again dominated by tip, tilt, and astigmatism. Subsequent iterations help to decrease the low frequencies even more down to a floor at $\sim$3~nm rms in all \cpup bins. After four iterations, the total amount of aberration in 1-15~\cpup decreases from 50-55~nm rms down to 9-12~nm rms, corresponding to a gain in wavefront error of  a factor of approximately five. High-spatial frequencies beyond 20~\cpup are mostly unaffected by the NCPA compensation, as expected.

Two important questions prior to on-sky tests pertained to whether or not ZELDA
would provide a measurable signal in the presence of ExAO residual phase errors, and to whether or not this signal would be accurate enough to enable a direct correction for the NCPA. The results in the right panel  of Fig.~\ref{fig:ncpa_loop_1} clearly show a positive answer to both questions. On sky, the OPD map at iteration \#0 is similar to the one on the internal source, with a noticeable astigmatism. Interestingly, the level of aberration measured on sky at iteration \#0 is lower than on the internal source: 35-40~nm rms versus 50-55~nm rms in 1-15~\cpup, respectively. This is partly due to the fact that on sky the tip-tilt at the beginning was compensated manually when centering the PSF on the ZELDA mask, and is also partly due to a loss in sensitivity of the sensor on sky; this latter is explored in Sect.~\ref{sec:zelda_sensitivity}. In terms of convergence of the NCPA loop, it seems that only two iterations are required to reach a plateau in the low spatial frequencies, resulting in a final value of 16~nm rms in 0-15~\cpup.

Overall we conclude that ZELDA on-sky measurements and NCPA compensation in closed loop are possible. However, there seems to be differences in the values measured between the internal source and on sky, which makes the absolute values of the $\sigma(f)$ on sky difficult to interpret.

\subsection{ZELDA sensitivity with atmospheric residuals}
\label{sec:zelda_sensitivity}

\begin{figure}
    \centering 
    \includegraphics[width=0.5\textwidth]{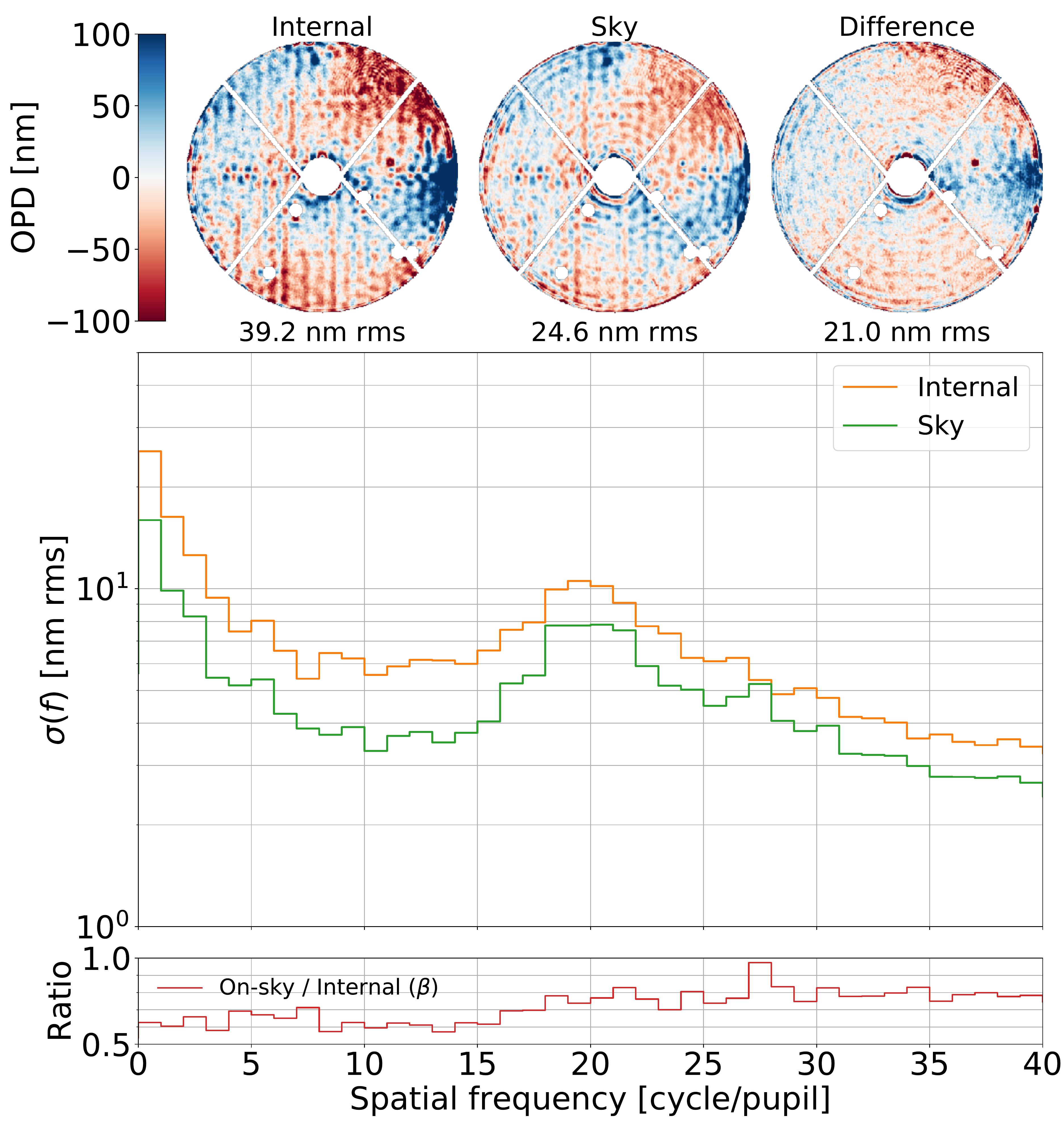}
    \caption{Comparison of the NCPA calibration performed on the internal source (left, test ZT06) and on sky (center, test ZT05). The top row shows the two OPD maps with the amount of aberration integrated in the range 1-15~\cpup. The main plot shows the $\sigma(f)$ of the two OPD maps, and the bottom plot shows the ratio of the on-sky and internal measurements. In this test, NCPA was first measured on sky, and subsequently the instrument was switched to internal source without changing anything other than the source-selection mirror before another NCPA measurement was acquired. The tip, tilt, and focus have been subtracted (see Sect.~\ref{sec:zelda_sensitivity}), which explains why the amount of integrated aberration is smaller than in Fig.~\ref{fig:ncpa_loop_1}).}
    \label{fig:ncpa_internal_sky}
\end{figure}

We investigated the apparent discrepancy between the internal and the on-sky measurement with a dedicated test performed on 2018-04-03. In this test, we first performed a ZELDA on-sky measurement on 1~Pup (test ZT05), then switched to the internal source without changing anything else in the instrument setup and finally made a new ZELDA measurement (test ZT06). Due to some time lost on switching to the internal source and on setting up the new ZELDA images, the internal measurement was only acquired $\sim$30 min after the on-sky one. 

\begin{figure}
    \centering 
    \includegraphics[width=0.5\textwidth]{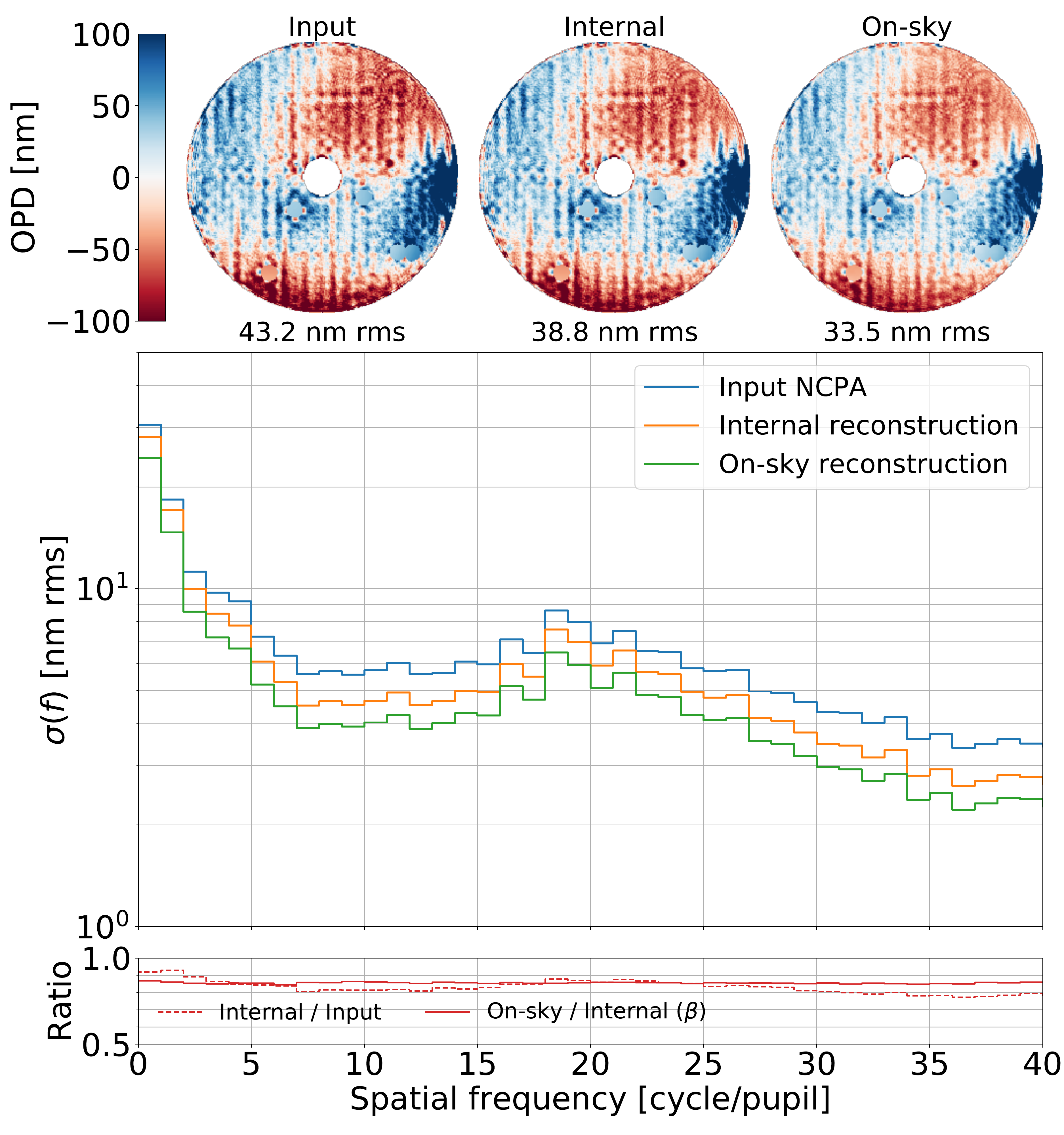}
    \caption{Numerical simulation of ZELDA wavefront reconstruction on the internal source and on sky, based on wavefront measurements from ZELDA and reconstructed ExAO residuals. The top row shows the OPD maps corresponding to the input NCPA injected into the simulation (left), the ZELDA-reconstructed NCPA on the internal source (center), and the ZELDA-reconstructed NCPA on sky (right). The amount of aberration integrated in the range 1-15~\cpup is reported below each OPD map. The main plot shows the $\sigma(f)$ of the three OPD maps, and the bottom plot shows the ratio of the simulation input and the internal simulated reconstruction and of the on-sky and internal simulated reconstructions. The on-sky simulation is based on 30~s of SAXO real-time telemetry acquired during our tests in Paranal (see Sect.~\ref{sec:zelda_sensitivity} and Appendix~\ref{sec:phase_screen_reconstruction}). The tip, tilt, and focus have been subtracted to match the analysis of the real data shown in Fig.~\ref{fig:ncpa_internal_sky}.}
    \label{fig:ncpa_internal_sky_simu}
\end{figure}

The results are presented in Fig.~\ref{fig:ncpa_internal_sky}. In this plot the tip, tilt, and defocus have been subtracted from the internal and on-sky OPD maps. For the tip-tilt, the DTTS ensures a repositioning accuracy of the order of 0.5~mas rms \citep{Sauvage2016} in a given setup. In these data, we measure differential tip and tilt of 1.43~mas and 4.15~mas, respectively. This is significantly larger than the specification, however here we are repositioning the PSF with the DTTS in two different setups: on sky with the full VLT pupil and on the internal source with a circular, nonobstructed pupil. The observed difference is therefore plausible, although more systematic tests would be required to confirm that the repositioning accuracy is slightly decreased between the two configurations. For the defocus, the observed difference of 14~nm rms is less easily explained. Apart from the presence of central obscuration and spiders, the other main difference between the internal source and the sky measurements is the illumination of the pupil, which is perfectly uniform on sky but Gaussian on the single-mode fiber of the internal source. One possibility is that the difference in illumination, in the presence of phase aberrations, propagates as a slight defocus in the Shack-Hartmann (SH) reconstruction. We tested this hypothesis using a simplified SH simulation using OOMAO \citep{Conan2014} but could not reproduce the observed defocus. At this stage the origin of this defocus is not precisely understood and further tests would be needed.

Once the tip, tilt, and focus are removed, the difference in sensitivity of ZELDA on sky appears very clearly in the $\sigma(f)$ plot, with a ratio of a factor $\beta = 0.64$ on average between the two over spatial frequencies up to 15~\cpup. The fact that the OPD maps are spatially almost identical but the $\sigma(f)$ values are at a decreased level indicates a change in sensitivity of ZELDA on sky, likely due to the difference in Strehl ratio between the two configurations. Beyond 15~\cpup, the ratio goes up to $\sim$0.8, possibly indicating a different regime at different spatial frequencies. From here on we use a fixed value for $\beta$ but keep in mind that the value could in fact vary as a function of the spatial frequency. We see further evidence of this possibility in Sect.~\ref{sec:analysis_limitations}.

In an attempt to explain the loss of sensitivity, we performed a numerical simulation of a ZELDA NCPA reconstruction based on realistic data acquired with the instrument. The simulation is designed to model measurements taken either on the internal source or on sky. For the on-sky simulation, we use reconstructed ExAO residuals computed from SAXO real-time telemetry data over 30~s (41\,400 phase screens; see Appendix~\ref{sec:phase_screen_reconstruction}). The average residual wavefront error of the generated phase screens is $95.9 \pm 12.4$~nm~rms, corresponding to a Strehl of $87\% \pm 3\%$ at 1.6~\mic, which is a good match to the observing conditions encountered during our tests in Paranal and the typical SAXO performance. The input static NCPA pattern is based on the ZELDA measurement performed in test ZT01 on the internal source and scaled to exactly 55 nm rms between 1 and 15~\cpup for the purpose of the simulation. This value corresponds to the one typically observed in SPHERE after the daily calibrations.

The results of the simulation are presented in Fig.~\ref{fig:ncpa_internal_sky_simu}. In this figure, the residual tip, tilt, and defocus have been removed for a fair comparison to Fig.~\ref{fig:ncpa_internal_sky}. We first note that the internal reconstruction does not match exactly the input NCPA, with a difference of a factor $\sim$0.8. This means that the internal reconstruction typically underestimates the NCPA by $\sim$20\%. We also clearly identify a difference in sensitivity between the internal source and on sky in this simulation and we measure a factor $\beta = 0.86$ of attenuation between the reconstructions simulated on sky and on the internal source for spatial frequencies up to 20~\cpup. This is higher than the factor 0.64 that was identified in the real data in Fig.~\ref{fig:ncpa_internal_sky}, and the difference in sensitivity appears much flatter than in the data, with a factor that remains the same beyond 20~\cpup. We also note a marginal variation of the sensitivity as a function of spatial frequency.

The differences between the internal source and on-sky results can only be attributed to the presence of two factors: the telescope optics and the atmospheric turbulence. Our numerical simulations already take into account the presence of the telescope optics by including the pupil illumination from pupil data measurements, and its effect is estimated to be negligible. The atmospheric turbulence residuals induce wavefront errors and possible cross-talk terms with the existing NCPA, which impact the Strehl ratio on sky. We therefore conclude that the value of the $\beta$ factor is primarily driven by the Strehl ratio.

The fact that the $\beta$ factor is not exactly identical in the simulation and on sky can likely be attributed to the variable observing conditions since the SAXO telemetry data were acquired almost 30~min after the ZELDA data. As a consistency check, we measured the Strehl ratio of an off-axis PSF acquired within $\sim$5~min of the SPARTA telemetry data to be 79\%. This value agrees reasonably well with the 82\% Strehl ratio expected for the $\sim$96~nm~rms of reconstructed residuals combined with $\sim$60~nm~rms of quasi-static NCPA. Multiplying the reconstructed ExAO residuals by a correction factor, we also determined that a Strehl ratio of $\sim$70\% is necessary to match the observed $\beta = 0.64$ attenuation factor. In conclusion we consider that the loss in sensitivity on sky is well understood and is directly related to the Strehl ratio of the observations. New ZELDA measurements obtained in parallel with SPARTA telemetry should hopefully enable a conclusion on this matter in the future.

\subsection{Influence of exposure time}
\label{sec:influence_integration_time}

\begin{figure}
    \centering 
    \includegraphics[width=0.5\textwidth]{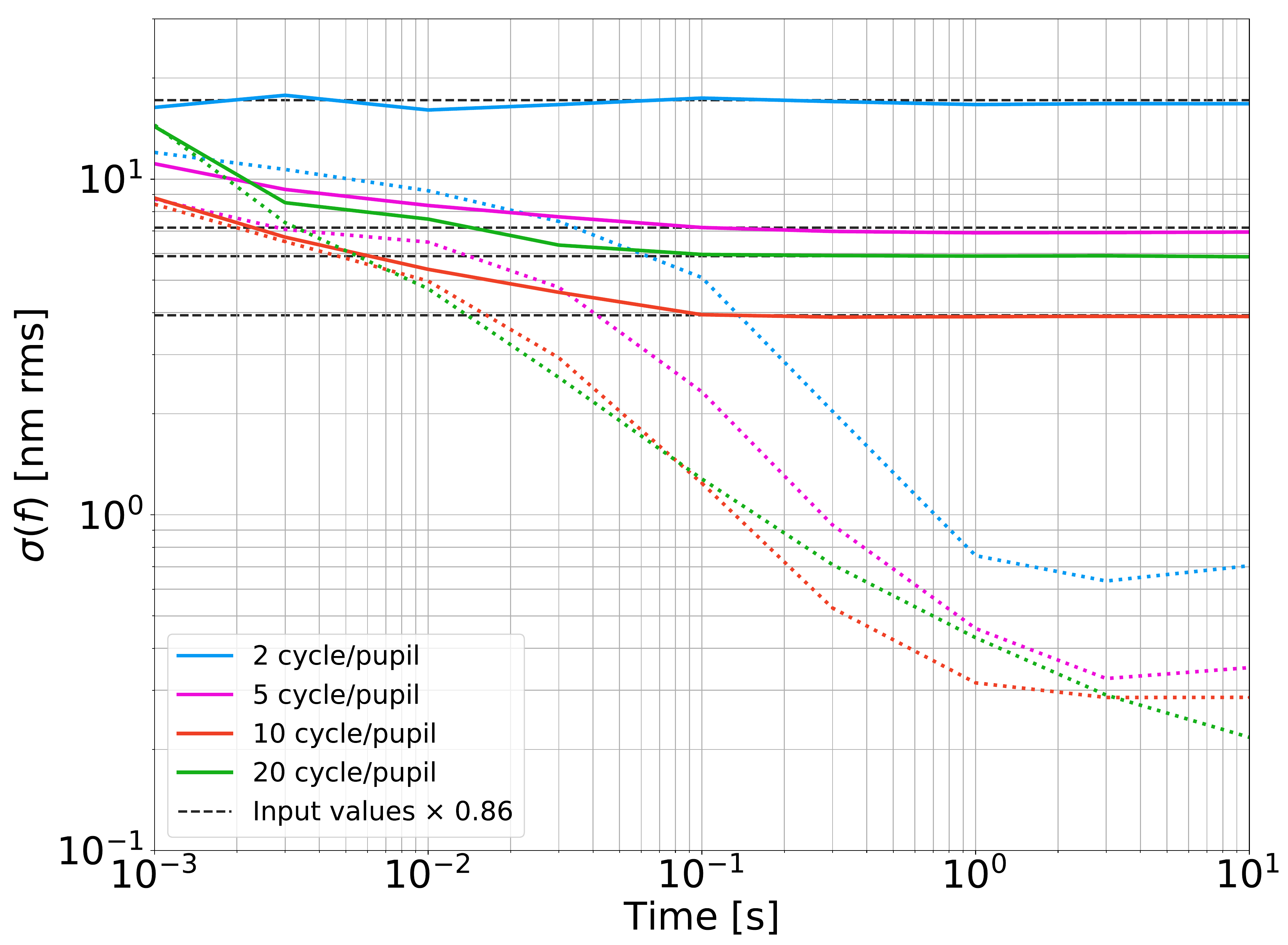}
    \caption{Influence of the integration time on the ZELDA reconstruction for 55 nm rms of NCPA and turbulence reconstructed from SAXO telemetry (plain lines), plotted at four different spatial frequencies. The dotted curves show the result of the reconstruction on pure atmospheric residuals, i.e., when setting the amount of NCPA to zero. The input values are corrected for the $\beta = 0.86$ sensitivity factor determined from simulation in Sect.~\ref{sec:zelda_sensitivity}.}
    \label{fig:zelda_sensitivity_integration_time}
\end{figure}

\begin{table*}
    \caption[]{Coronagraphic imaging log}
    \label{tab:coro_log}
    \centering
    \begin{tabular}{lccccccccccc}
    \hline\hline
    Test name  & Date         & UTC time  & Source       & DIMM seeing     & $\tau_{0}$    & \multicolumn{2}{c}{NCPA}     & DIT & NDIT & NEXP & Collapse \\
               &              &           &              &                 &               & calib.           & iter.     &     &      &      &          \\
               &              &           &              & (\as)           & (ms)          &                  &           & (s) &      &      &          \\ 
    \hline
    CT01\_ref  & 2018-04-01   & 19:30     & Internal     &                 &               & ZT01             & 0         & 1   & 60   & 1    & yes      \\
    CT01       & ...          & ...       & ...          &                 &               & ...              & 3         & 1   & 60   & 1    & yes      \\
    CT02\_ref  & 2018-04-02   & 03:10     & $\alpha$ Crt & 0.70 $\pm$ 0.09 & 5.1 $\pm$ 0.5 & ZT02             & 0         & 7   & 10   & 4    & yes      \\
    CT02       & ...          & ...       & ...          & ...             & ...           & ...              & 3         & 7   & 10   & 4    & yes      \\
    \hline
    CT03\_ref  & 2018-04-03   & 17:45     & Internal     &                 &               & ZT03             & 0         & 10  & 12   & 1    & yes      \\
    CT03       & ...          & ...       & ...          &                 &               & ...              & 3         & 10  & 12   & 1    & yes      \\
    CT04\_ref  & 2018-04-04   & 00:39     & 1 Pup        & 0.71 $\pm$ 0.05 & 4.7 $\pm$ 0.3 & ZT04             & 0         & 5   & 12   & 4    & no       \\
    CT04       & ...          & ...       & ...          & ...             & ...           & ...              & 3         & 5   & 12   & 4    & no       \\
    \hline
    \end{tabular} 
    \tablefoot{Missing value indicates no applicable value for the test and ``...'' indicates a value identical to the previous line.}
\end{table*}

To make sure that our on-sky NCPA reconstruction with ZELDA is not limited by the residual atmospheric turbulence downstream of the ExAO system in typical observing conditions (0.7\as seeing, 5~ms $\tau_0$, bright star) and AO regime (1.3 kHz loop speed), we also explored the effect of integration time for the on-sky NCPA reconstruction in numerical simulations. We used the same static NCPA pattern as in Fig.~\ref{fig:ncpa_internal_sky_simu} and simulated reconstructions assuming exposure times from 0.001~s to 10~s. The results are showed in Fig.~\ref{fig:zelda_sensitivity_integration_time}. In this plot we represent the $\sigma(f)$ values of the reconstructed NCPA for four different values of $f$ as a function of integration time as well as a reconstruction on pure atmospheric residuals, that is when setting the amount of NCPA to zero, to show the impact of these residuals on the reconstruction.

At very small integration times, the static NCPA and the turbulent parts are both an issue and contribute equally in the reconstructed wavefront. Beyond 0.1~s, the curves drop rapidly, which means that the atmospheric residuals, that is dynamic contributions to the speckle field, become negligible to the standard deviation of the wavefront and hence in this regime it is critical to specifically address the NCPA terms as they dominate. We know from Fig.~\ref{fig:ncpa_loop_1} that the level of $\sigma(f)$ will drop down to $\sim$4~nm~rms in the 1-15~\cpup range once the NCPA is compensated for. Setting up a DIT longer than $\sim$0.1~s is therefore required to force the dynamic terms to fall below this threshold value and measure the NCPA accurately. Finally, we can safely conclude that we can perform NCPA compensation with ZELDA on bright stars with 8-10m telescopes equipped with ExAO using short exposure times.

%--------------------------------------------------------------------------------------------------------------------
\section{Coronagraphic performance}
\label{sec:coronagraphic_performance}

\subsection{Description of the tests}
\label{sec:decription_tests_coro}

Beyond the pure on-sky NCPA compensation, the goal of our tests was to demonstrate a quantitative contrast gain in raw coronagraphic images once the compensation is applied. For this purpose, we acquired coronagraphic data following the ZT01, ZT02, ZT03, and ZT04 tests presented in Table~\ref{tab:ncpa_loop_log}. The complete data logs for these tests are provided in Table~\ref{tab:coro_log}.

We used the SPHERE apodized-pupil Lyot coronagraph \citep[APLC;][]{Soummer2005,Carbillet2011,Guerri2011} and all data were acquired with the IRDIS instrument \citep{Dohlen2008} in dual-band imaging \citep[DBI;][]{Vigan2010} mode. To present the results we focus on the images in the H2 filter of the H23 filter pair, at a wavelength of 1.593~\mic and with a width of 52~nm. For each coronagraphic test, we acquired a set of coronagraphic images on-axis, followed by an off-axis PSF taken with a neutral density filter, and finally an instrumental background obtained by either switching the lamp off (internal source) or applying a 30\as telescope offset (on sky). The DITs were manually adjusted to avoid any saturation of the detector. The centering of the PSF on the coronagraph mask was manually performed so as to minimize any visible residual tip-tilt. 

On the internal source, we acquired two short-exposure coronagraphic images ($\sim$1~min): one with the HOWFS reference slopes corresponding to iteration \#0 in the NCPA compensation loop (referred to as ``before compensation'' hereafter), and one with reference slopes corresponding to iteration \#3 (``after compensation'' hereafter). A similar procedure was applied on sky but the before and after compensation images were interleaved over several minutes to better sample the variable observing conditions. On 2018-04-03 all the DIT were saved independently, while on 2018-04-01 a setup error ended up saving only the co-added image of 10 NDIT. On 2018-04-03 we also saved SAXO real-time telemetry data in parallel with most of the coronagraphic data acquisition to reconstruct the ExAO residual wavefront errors a posteriori.

\subsection{Results}
\label{sec:results_coro}

\begin{figure*}
    \centering 
    \includegraphics[width=0.49\textwidth]{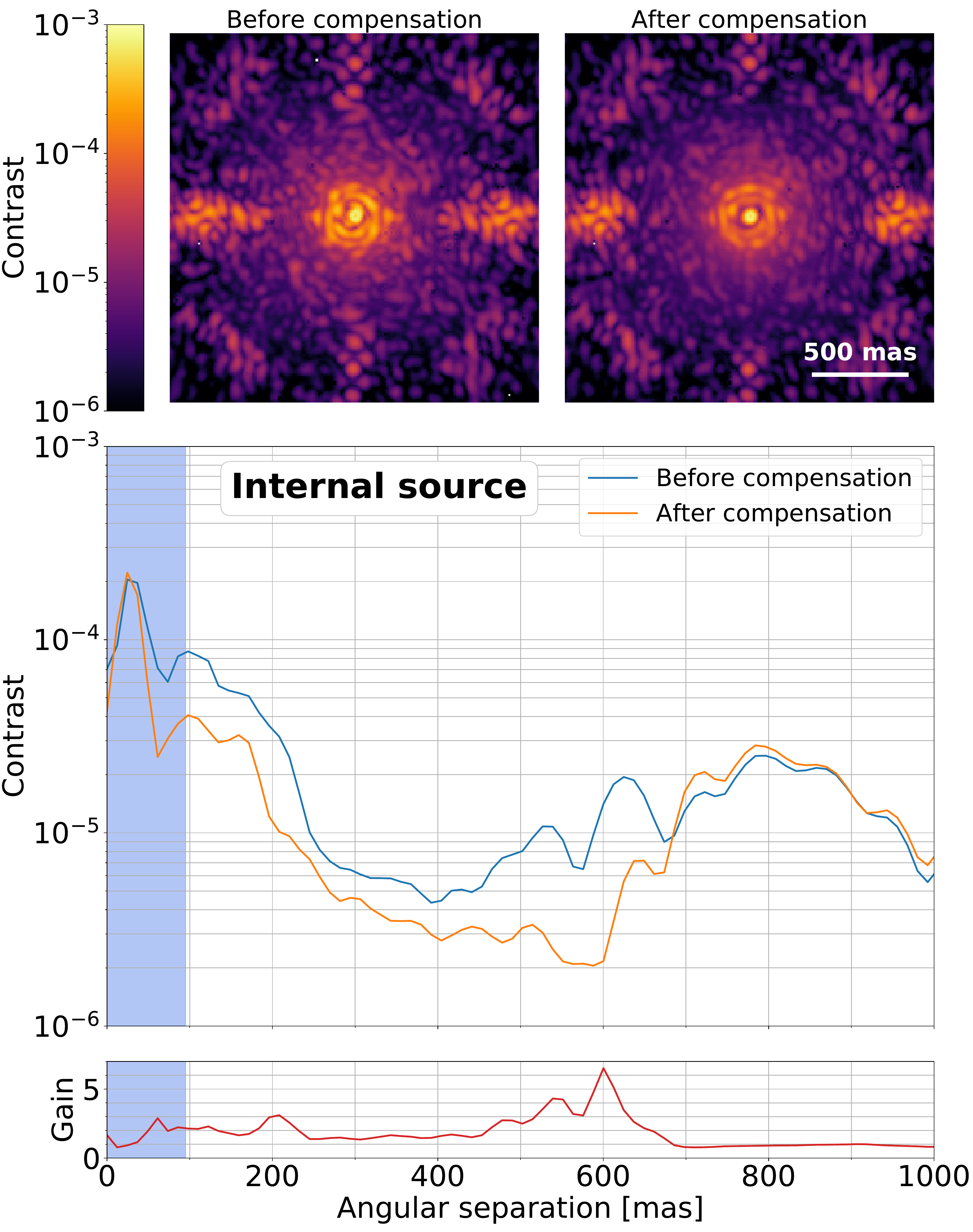}
    \includegraphics[width=0.49\textwidth]{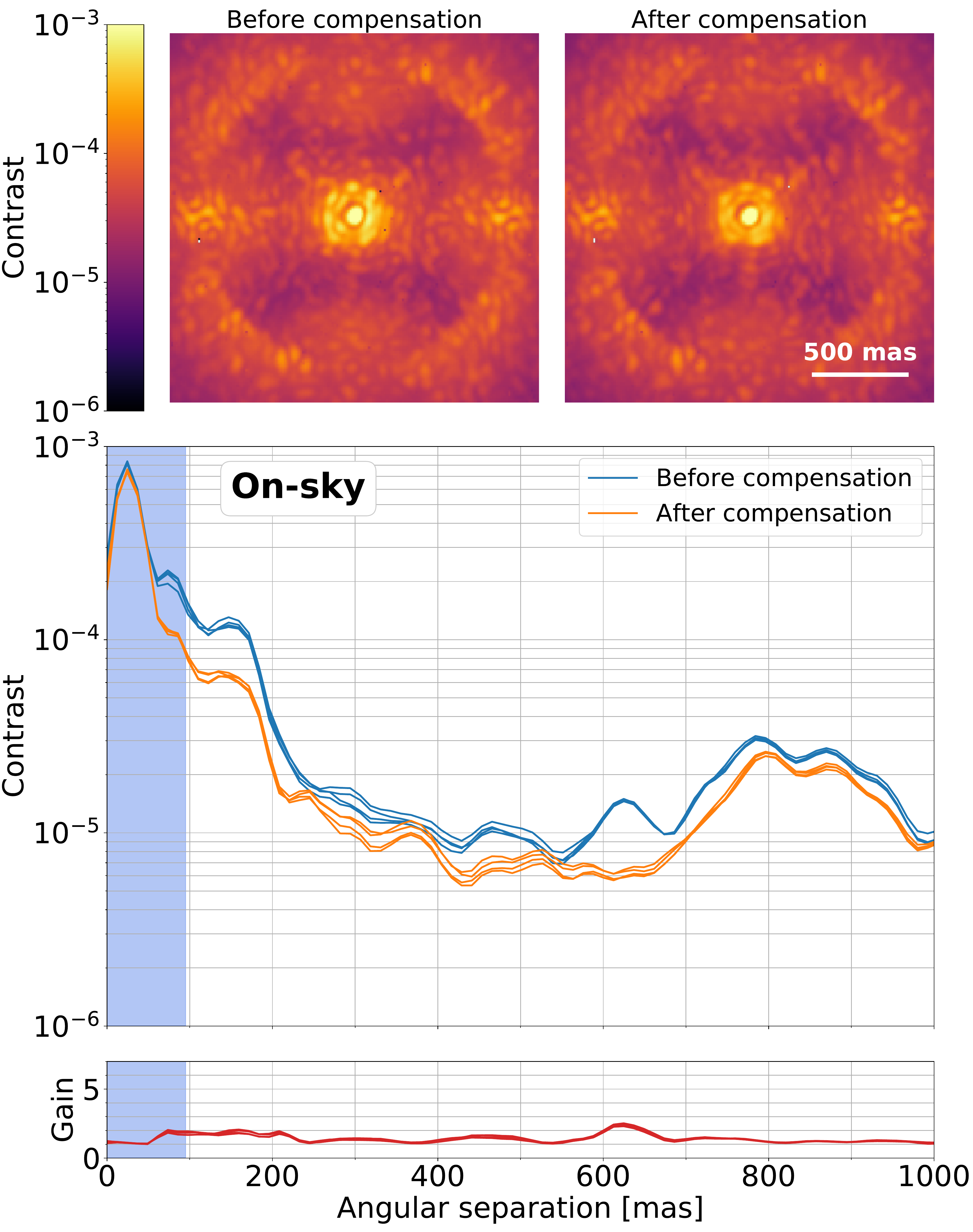}
    \caption{Coronagraphic images and profiles obtained on 2018-04-01 on the internal source (left, test CT01) and on sky (right, test CT02). The top row shows the coronagraphic images before and after NCPA compensation, the main plots show azimuthal standard deviation profiles normalized to the off-axis PSF, and the bottom plots show the contrast gain after NCPA compensation. For on-sky measurements, a profile is shown for each individual coronagraphic image. The blue shadowed region shows the region masked by the focal-plane mask of the APLC.}
    \label{fig:coro_profiles_1}
\end{figure*}

The results on 2018-04-01 on the internal source and on sky are presented in Fig.~\ref{fig:coro_profiles_1}. On the internal source, there is a visible gain from the NCPA correction in the images. The gain is particularly visible on the symmetry of the residuals at the edge of the focal-plane mask, as well as on the vertical and horizontal structures created by the HODM actuators on the edge of the control region. These structures are not corrected up to the AO cutoff frequency because only 700 KL modes are kept in the NCPA compensation (see Sect.~\ref{sec:decription_tests_ncpa}). The gain is visible on the contrast curves, with the largest gain being of a factor of six at $\sim$600~mas, which corresponds to the location of the structures mentioned above. At very small separation (100-300~mas), the gain is limited to a factor of two or three.

On sky the coronagraphic images are dominated by the ExAO residuals in the control region, with a significant contribution from the aliasing term (Cantalloube et al. submitted to ESO Messenger) that creates the bright horizontal and vertical lobes starting at the cutoff frequency and decreasing towards the center leading visually to a dark cross in the AO-controlled region. In these observations, the spatial filter of the Shack-Hartmann wavefront sensor was set to its \texttt{MEDIUM} position (1.5~\lsd), which provides a moderate attenuation of the aliasing. The \texttt{SMALL} position (1.3~\lsd), which provides a much stronger attenuation of the aliasing, could not be used given the observing conditions on that date. The pattern of quasi-static speckles that was visible on the internal source is still visible in most of the on-sky image. We observe a small contrast gain from 100~mas up to 700~mas after NCPA compensation, but this gain is limited to a factor of 2.5 at most. Similarly to the internal source data, the highest gain is observed at a separation of 600~mas.

\subsection{Analysis of the limitations with coronagraphic image reconstruction}
\label{sec:analysis_limitations}

\begin{figure*}
    \centering 
    \includegraphics[width=0.49\textwidth]{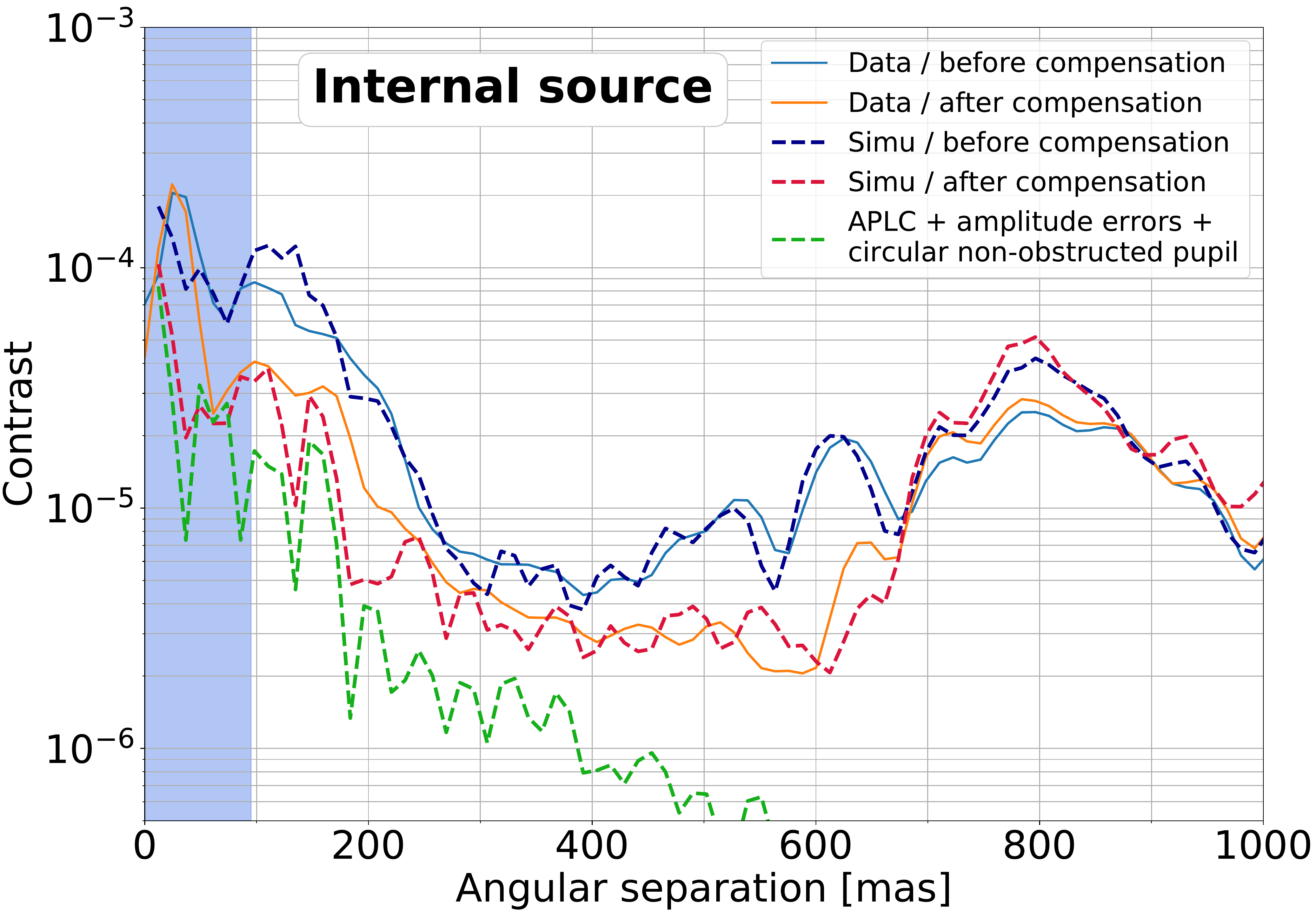}
    \includegraphics[width=0.49\textwidth]{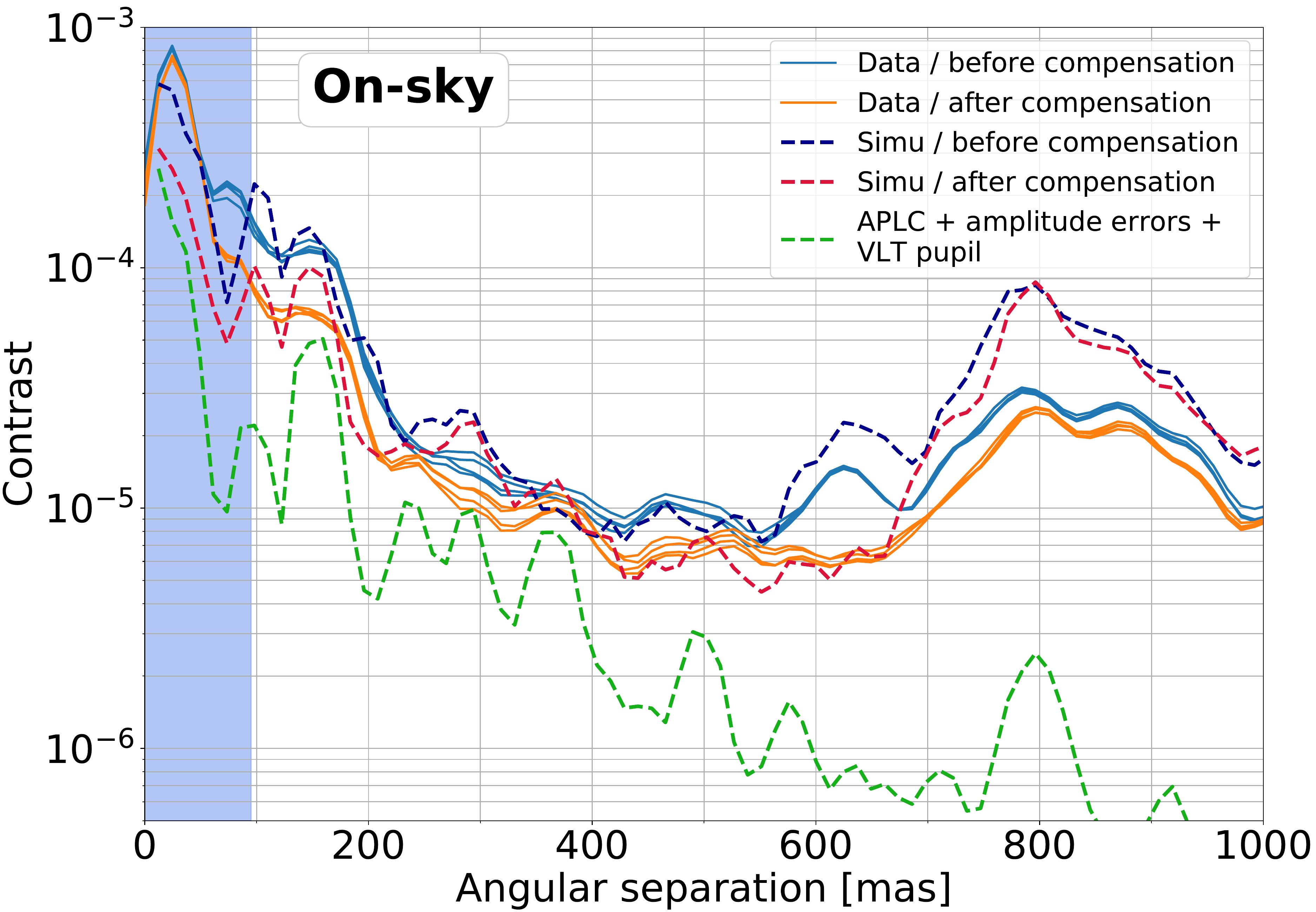}
    \caption{Comparison of the coronagraphic profiles acquired on 2018-04-01 on the internal source (left; test CT01) and on sky (right; test CT02) with coronagraphic image reconstruction based on realistic inputs (see Appendix~\ref{sec:coronagraphic_image_reconstruction}). The theoretical performance of the APLC in the presence of amplitude errors is also plotted (dashed green curve). For the on-sky data, the image reconstruction includes ExAO residual wavefront errors based on SAXO real-time telemetry (see Appendix~\ref{sec:phase_screen_reconstruction}). The blue shadowed region shows the region masked by the focal-plane mask of the APLC.}
    \label{fig:coro_profiles_simu_1}
\end{figure*}

To understand the observed results on the internal source and on sky we performed coronagraphic image reconstruction based on realistic inputs fed into a model of the instrument. The details of the reconstruction are provided in Appendix~\ref{sec:coronagraphic_image_reconstruction}. 

The comparison between the data and the reconstruction is presented in Fig.~\ref{fig:coro_profiles_simu_1}. On the internal source, the reconstruction provides an excellent match to the data at low spatial frequencies ($\leq650$~mas), provided that a defocus term of -40~nm~rms is added in the model. This term originates from the fact that the ZELDA mask and the APLC focal-plane masks are mounted in two different positions of the same filter wheel. There is therefore no reason for the best centering and focus to be exactly identical between the two. To find the best value, we generated a family of reconstructions with an increasing defocus term from -100 to +100~nm~rms in steps of 10~nm~rms and compared the resulting contrast curves with the data. The best fit was obtained for $\sim$0~nm of defocus for the data before NCPA compensation and -40~nm~rms for the data after NCPA compensation. Physically, the $\sim$0~nm defocus before compensation is expected because it corresponds to the HO-WFS reference slopes optimized for the usual coronagraph during the daily calibrations. After compensation, the focus is, on the contrary, optimized for the ZELDA mask, hence the need to add an appropriate defocus term.

For the image reconstruction, we also multiply the ZELDA OPD map with a factor to compensate for the loss of sensitivity of the ZELDA reconstruction. As demonstrated in the simulation in Fig.~\ref{fig:ncpa_internal_sky_simu}, the value of this loss factor is of the order of $0.80$ on the internal source, so we use $1/0.8$ as the correction value in the reconstruction of the internal source data.

The data and reconstruction before and after NCPA compensation match very well at separations below 700~mas. Around the AO cutoff frequency the simulation predicts a level slightly higher than what is actually observed, which could be related to a variation of the sensitivity factor as a function of spatial frequency (see Sect.~\ref{sec:zelda_sensitivity}). After compensation, close to the focal-plane mask, two bumps at $\sim$100~mas and $\sim$175~mas are clearly identifiable in both curves. We also compare these results with the contrast curve simulated for the APLC with a circular nonobstructed pupil and in the presence of amplitude errors. At the smallest separations, the raw contrast after compensation of the NCPA on the internal source is very close to the theoretical limit of the APLC when taking into account the amplitude aberrations. This means that on the internal source we are likely limited by the coronagraph design at separations <200~mas, and not by our capacity to calibrate and compensate for NCPA, although there seems to remain a very small amount of low-order NCPA. Beyond 200~mas, the profiles of the theoretical APLC and the data depart from each other, which means that the mid-frequencies ($\sim$5-15~\cpup) are not completely compensated for. Additional work is still needed to understand why the NCPA compensation does not allow us to  further decrease the mid-spatial frequency errors. One likely possibility is the presence of several dead actuators in the SPHERE HODM, and in particular a column of them located just on the outside on the right of the pupil. While the dead actuators are masked in Lyot stop plane to decrease their impact on coronagraphic imaging, their presence necessarily has an effect on what can be corrected. The column of bad actuators shows an extended impact on a significant portion of the pupil, which can be seen on the OPD maps at iterations \#2-4 in Fig.~\ref{fig:ncpa_loop_1}. Unfortunately their impact is difficult to model accurately because the phase error induced by the dead actuators is far beyond the linearity range of ZELDA.

On sky, the reconstruction also matches the data extremely well below $\sim$700~mas. However, at the level of the AO cutoff there is a sharp increase in the simulated curve that is not observed in the data. In this reconstruction, the OPD map measured with ZELDA is compensated assuming a sensitivity loss factor of $\beta = 0.64$ as determined from actual measurements in Sect.~\ref{sec:zelda_sensitivity}. Nevertheless, the $\beta$ factor represents the difference between the on-sky measurement and the internal source measurement, but we see in Sect.~\ref{sec:zelda_sensitivity} that the internal measurement is also attenuated by a factor 0.8 with respect to the real NCPA value. As a result, the on-sky ZELDA measurement is compensated by a factor $1/0.64 \times 1/0.8 = 1.95$ in the reconstruction. While this factor seems appropriate for the low-spatial frequencies, the value does not seem appropriate at the level of the AO cutoff frequency. This again points towards a value that would vary as a function of spatial frequency. For this first study, provided that the predictions of our simulations appear correct at small separation, we keep a fixed value for the attenuation factor. However, further study would be required to fully understand the behavior of the sensitivity loss factor.

Other than the atmospheric residuals, the main change on sky is the VLT pupil. The impact of this pupil is significant on the theoretical APLC raw contrast, with the apparition of a bright peak at 150~mas as well as secondary peaks in the 200-400~mas range. In the 100-400~mas range, the data after NCPA compensation are obviously limited by these peaks. The relatively small raw contrast gain after NCPA correction is therefore not induced by an improper NCPA compensation on sky, but simply by the absolute performance of the SPHERE APLC design \citep{Carbillet2011,Guerri2011}. Even before NCPA compensation, the data are also visibly limited by the coronagraph, although to a lesser extent, probably through an effect similar to pinned speckles \citep{Bloemhof2001,Sivaramakrishnan2002,Perrin2003}. In conclusion, we can verify that the NCPA compensation with ZELDA functions on sky, but we cannot evaluate its ultimate performance due to the current design of the SPHERE APLC.

%--------------------------------------------------------------------------------------------------------------------
\section{Discussion}
\label{sec:discussion}

Second-generation exoplanet imagers have already demonstrated great potential for studying close circumstellar environments, and for detecting new companions and helping to understand their physical properties. However, the current statistics derived from observations \citep[e.g.,][]{Nielsen2019arXiv} and the most recent planet population models \citep[e.g.,][]{Mordasini2017,Forgan2018} respectively show and predict that giant planets are scarce in the regime of mass and separation probed by current instruments. In other words, we need to probe closer to the stars and at deeper contrasts. 

One of the current limitations in VLT/SPHERE is the control and compensation for the NCPA. To address this issue, we proposed to use the ZELDA wavefront sensor \citepalias{N'Diaye2013,N'Diaye2016} to measure these aberrations down to the level of the coronagraphic focal-plane mask and compensate for them with the HODM. The results on the internal light source in the present work clearly demonstrate that a major gain in the amount of aberration between 1 and 15~\cpup can be obtained, with a decrease from $\sim$50~nm~rms down to less than 10~nm~rms. This compensation for NCPA naturally translates into a gain in raw contrast in the coronagraphic images, although our simulations based on coronagraphic image reconstruction show that the performance is not limited by the final amount of NCPA after compensation but by the design of the SPHERE APLC. We cannot therefore conclude easily on the ultimate contrast that could be reached thanks to NCPA compensation with ZELDA.

On sky, we demonstrated that ZELDA measurements are possible and that they enable closed-loop NCPA compensation in a fashion identical to that performed on the internal source. However, our data show an important loss in sensitivity of ZELDA, which cannot be explained solely by a loss in Strehl ratio. The sensitivity loss does not represent a limitation as long as the NCPA compensation is performed in closed loop. The compensation will potentially require more iterations to reach the same level of correction, even though in practice (Fig.~\ref{fig:ncpa_loop_1}) it seems that a plateau is reached in only a few iterations. In any case, we establish that the presence of ExAO residuals does not limit the accuracy of the ZELDA measurement for integration times beyond a few tenths of a second. A system implementing a continuous NCPA compensation loop would therefore have to run at a very minimum frequency of 10~Hz.

In terms of coronagraphic performance after NCPA compensation, we show that raw contrast gain is of the order of a factor of 2 to 2.5 at most within 100-700 mas. While this gain is modest, our coronagraphic image reconstructions clearly demonstrate that the SAXO AO system and APLC are performing close to their theoretical limits, in particular when looking at very small angular separation (<400~mas). This comes as a surprise because we anticipated that on sky the main limitation would be dominated by the ExAO residuals close to the optical axis. Our NCPA correction approach based on ZELDA would however offer greater gains on other imagers that are not as well calibrated, or for a coronagraph design with increased theoretical contrast at short separation.

\begin{figure}
    \centering 
    \includegraphics[width=0.49\textwidth]{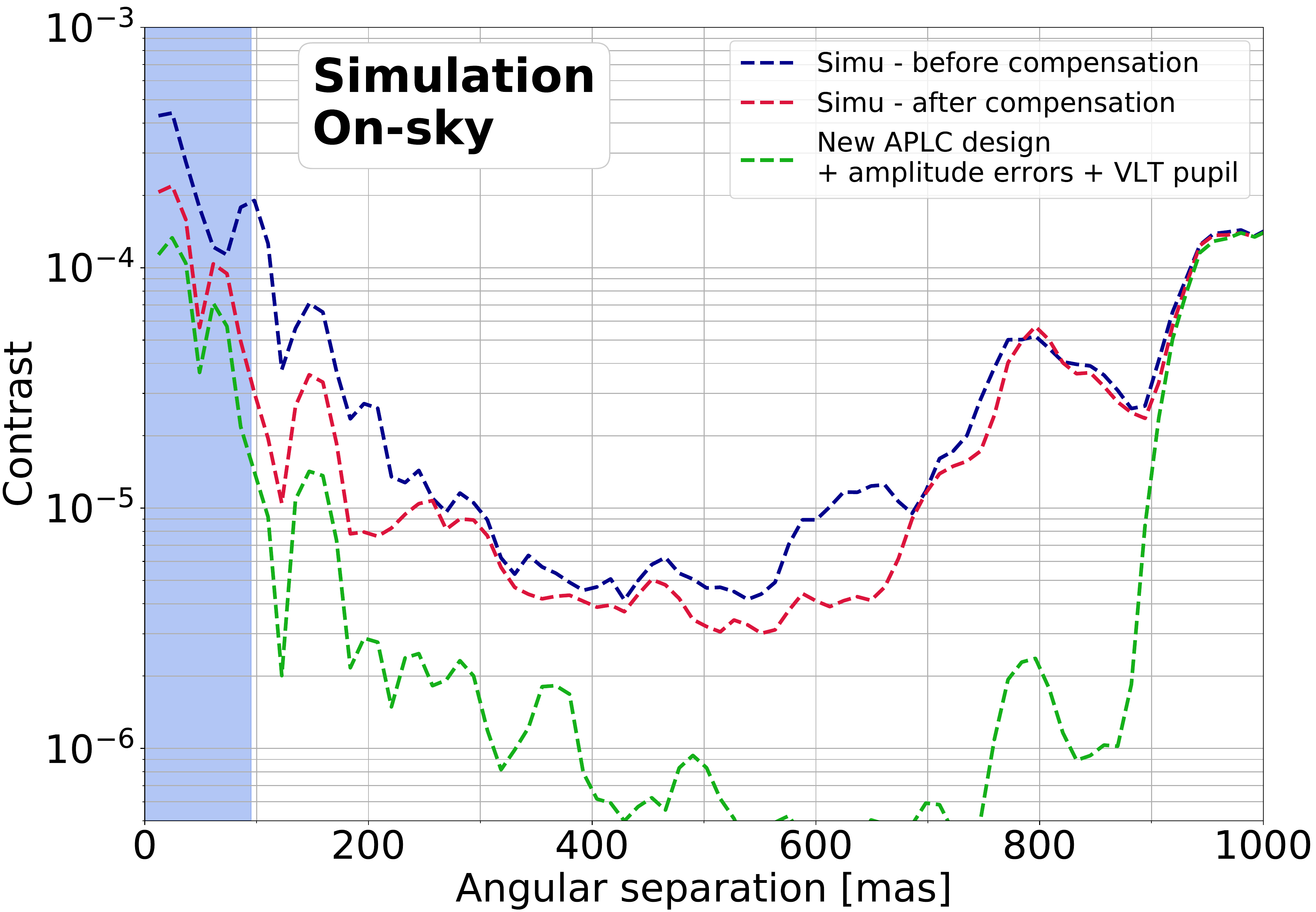}
    \caption{Expected raw-contrast performance on sky with a new apodizer design for the SPHERE APLC, before and after compensation of the NCPA. The design has been optimized to maximize the contrast in the 80-820~mas region \citep{N'Diaye2016b}. The theoretical performance of the APLC in the presence of amplitude errors is also plotted (dashed green curve). The coronagraphic image reconstruction includes ExAO residual wavefront errors reconstructed from SAXO real-time telemetry (see Appendix~\ref{sec:phase_screen_reconstruction}). The blue shadowed region shows the region masked by the focal-plane mask of the APLC.}
    \label{fig:new_aplc_design_coro_simu}
\end{figure}

Indeed, this result has strong implications in the context of discussions around upgrades for the SPHERE instrument \citep{Lovis2017,Vigan2018,Mouillet2018,Beuzit2019}. One of the main goals of such an upgrade would be to improve the sensitivity of the instrument at very small angular separation. With the present work we show that even without a heavy AO upgrade, a low-complexity coronagraph upgrade combined to a proper NCPA compensation scheme would potentially bring a significant contrast gain below 400~mas, i.e., in the region of interest for new planetary companions. To illustrate this, we designed an APLC based on the current focal-plane mask and Lyot stop but with a new apodizer that maximizes the contrast in the 80-820~mas region while having the same transmission as the current \texttt{APO1} apodizer \citep{N'Diaye2016b}. The outer limit of 820~mas has been chosen to cover the size of the AO control region in $H$-band (825~mas at 1.6~\mic). We injected the resulting apodizer in our coronagraphic image model and the results are shown in Fig.~\ref{fig:new_aplc_design_coro_simu}. Although the raw contrast appears still somewhat limited by the coronagraph at 150~mas, the raw contrast is now a factor of more than two lower than with the current APLC design. Quite importantly, the coronagraph is also no longer a visible limitation in the 200-400~mas range. Without implying that this specific design is the one that should be implemented in a SPHERE upgrade, it demonstrates that a simple change of apodizer could immediately bring a quantitative gain in contrast.

A better coronagraph design would however require a real NCPA control scheme that is fully part of the operational model of the SPHERE instrument. More generally, the implementation of a ZELDA-based NCPA compensation scheme in any exoplanet imager could follow two different scenarios.

The first one is to have the ZELDA device and the coronagraphic mask in the same filter wheel and to switch between wavefront sensing and science imaging. This scenario has mainly been explored in SPHERE and could be easily transposable to other existing instruments in which a ZELDA component can be added in the coronagraphic mask focal plane. In this scenario, one can calibrate the NCPA either (i) in daytime or during the telescope pointing, which is the most efficient in terms of telescope time, or (ii) directly on-sky on the science target before the science observations. However, this strategy relies on an intrinsic stability of the NCPA as a function of time and/or instrumental configuration. In the case of SPHERE, these stability aspects will be treated in a forthcoming paper. The constraints are even stronger in (ii), since this calibration requires some knowledge of the integration times and frame rates to use to reach sufficient accuracy in the determination of the NCPA. Another important aspect is the chromaticity of the NCPA and its correction. This aspect has begun to be addressed in \citetalias{N'Diaye2013} and \citetalias{N'Diaye2016} but a more thorough study is required to conclude on the effectiveness of the correction over a wide spectral band. This is highly relevant for exoplanet imagers with spectroscopic capabilities to study, for example, exoplanet atmospheres over a wide spectral window. During our latest observing run, suitable data have been acquired with SPHERE to the spectral behavior of the NCPA, which will be the subject of future work.

The second and ideal scenario would be a measurement and compensation done regularly in parallel with the science observations, for example performing ZELDA wavefront sensing in a spectral window located outside of the science wavelengths of interest. In the case of SPHERE, this approach is already in use by the DTTS, which picks up a small fraction of the light in $H$-band just before the coronagraph to measure and compensate for the residual pointing errors of the PSF. The DTTS could in principle be replaced by a ZELDA-based system to sense not only tip-tilt errors but also errors of a higher order.

The use of ZELDA for NCPA compensation is a promising approach for SPHERE, and could be implemented easily in other existing exoplanet imagers. Its application is already envisioned in upcoming facilities on the ground and in space to compensate for NCPA but also for low-order wavefront-sensing aspects. In particular, the approach analyzed here has been adopted as the baseline for NCPA compensation of ELT/HARMONI in its high-contrast imaging mode \citep{Carlotti2018} and for the low-order wavefront sensor of WFIRST/CGI \citep{Zhao2014}. The maturity of ZELDA gained in these high-visibility instruments will pave the way for its use in space observatories with exo-Earth imaging capabilities such as LUVOIR or Habex.

\begin{acknowledgements}
The authors are extremely grateful to ESO for granting them VLT technical night-time for these tests and to the Paranal observatory staff for their support. They also thank the referee for his insightful comments that greatly helped to improve the paper. This project has received funding from the European Research Council (ERC) under the European Union's Horizon 2020 research and innovation programme (grant agreement No. 757561). AV and MN acknowledge partial support from the French national programme ``\emph{Action Sp\'ecifique Haute R\'esolution Angulaire}'' (ASHRA). SPHERE is an instrument designed and built by a consortium consisting of IPAG (Grenoble, France), MPIA (Heidelberg, Germany), LAM (Marseille, France), LESIA (Paris, France), Laboratoire Lagrange (Nice, France), INAF - Osservatorio di Padova (Italy), Observatoire de Gen\`eve (Switzerland), ETH Z\"urich (Switzerland), NOVA (Netherlands), ONERA (France) and ASTRON (Netherlands) in collaboration with ESO. SPHERE was funded by ESO, with additional contributions from CNRS (France), MPIA (Germany), INAF (Italy), FINES (Switzerland) and NOVA (Netherlands). SPHERE also received funding from the European Commission Sixth and Seventh Framework Programmes as part of the Optical Infrared Coordination Network for Astronomy (OPTICON) under grant number RII3-Ct-2004-001566 for FP6 (2004-2008), grant number 226604 for FP7 (2009-2012) and grant number 312430 for FP7 (2013-2016).
\end{acknowledgements}

\bibliographystyle{aa}
\bibliography{paper}

\appendix

\section{Results obtained on 2018-04-03}
\label{sec:additional_results}

\begin{figure*}
    \centering 
    \includegraphics[width=0.49\textwidth]{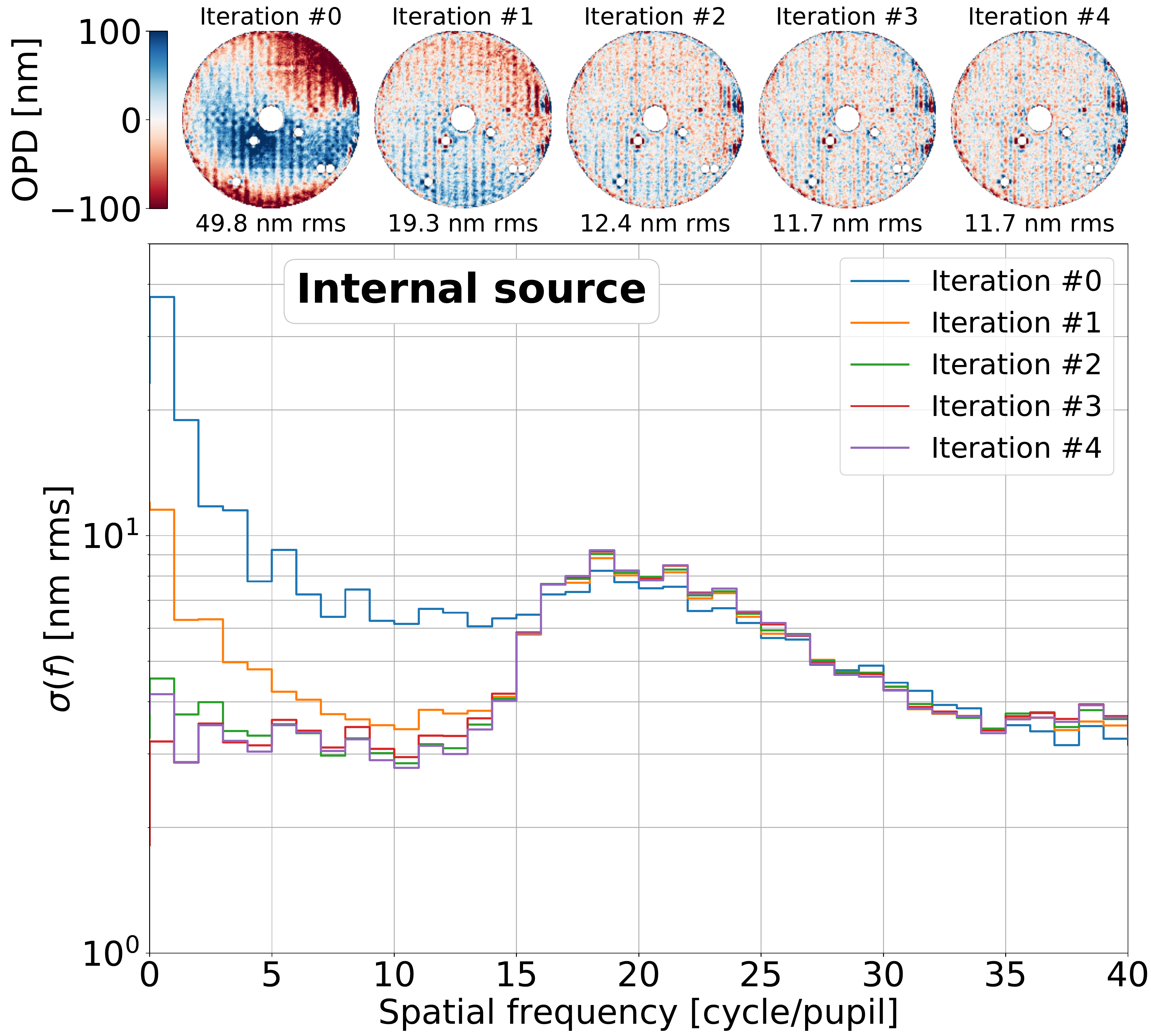}
    \includegraphics[width=0.49\textwidth]{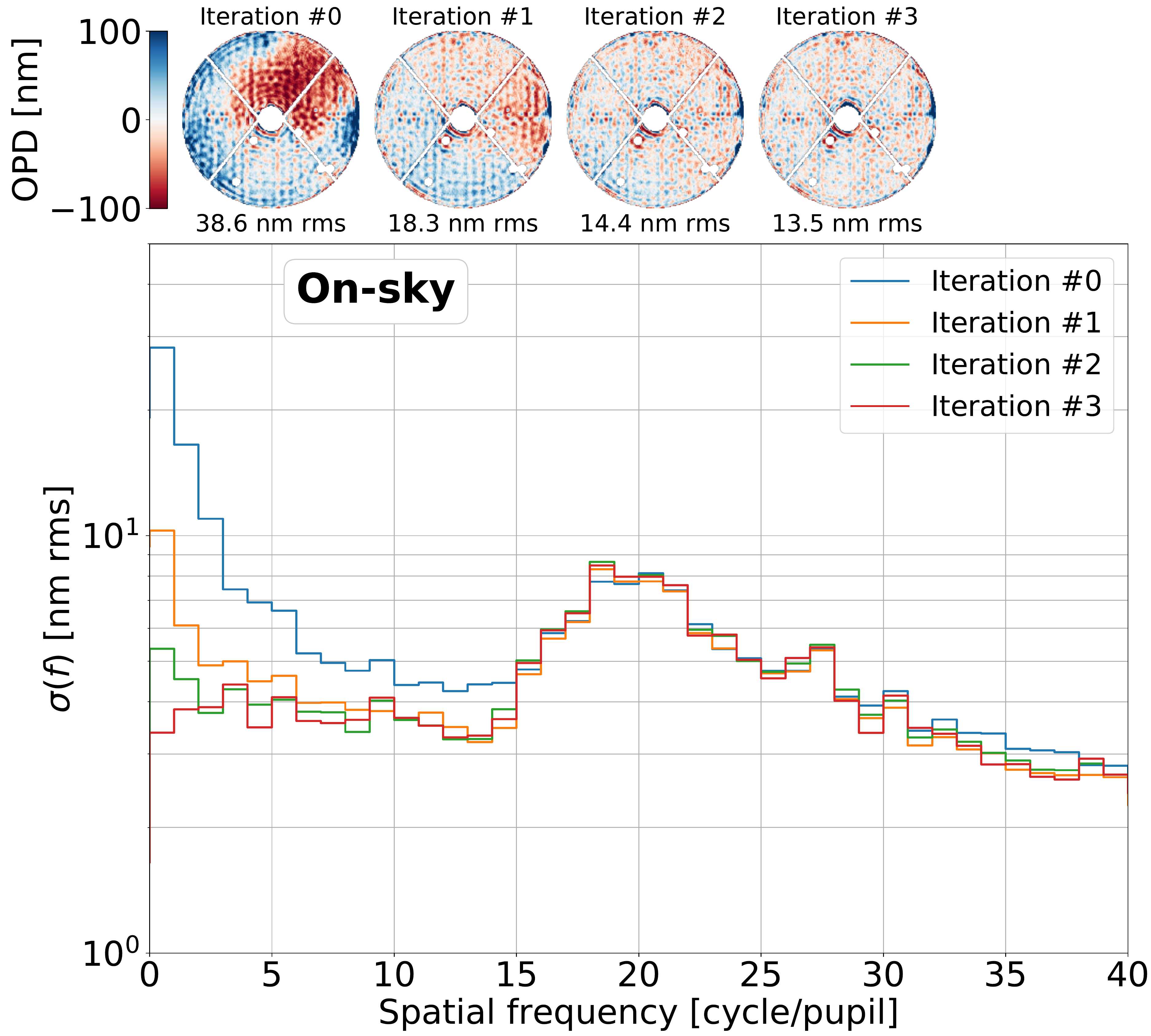}
    \caption{NCPA compensation loop results on 2018-04-03 on the internal source (left, test ZT03) and on sky (right, test ZT04). The top row shows the ZELDA OPD maps at the start (iteration \#0) and at the subsequent iterations. The value reported below each OPD map corresponds to the amount of aberrations integrated in the range 1-15~\cpup. The bottom plot shows the $\sigma(f)$ of the OPD map at each iteration (see Sect.~\ref{sec:results_ncpa} for the $\sigma(f)$ definition). In addition to the dead actuators and central part of the pupil that are always masked, the spiders are also masked for the on-sky test.}
    \label{fig:ncpa_loop_2}
\end{figure*}

\begin{figure*}
    \centering 
    \includegraphics[width=0.49\textwidth]{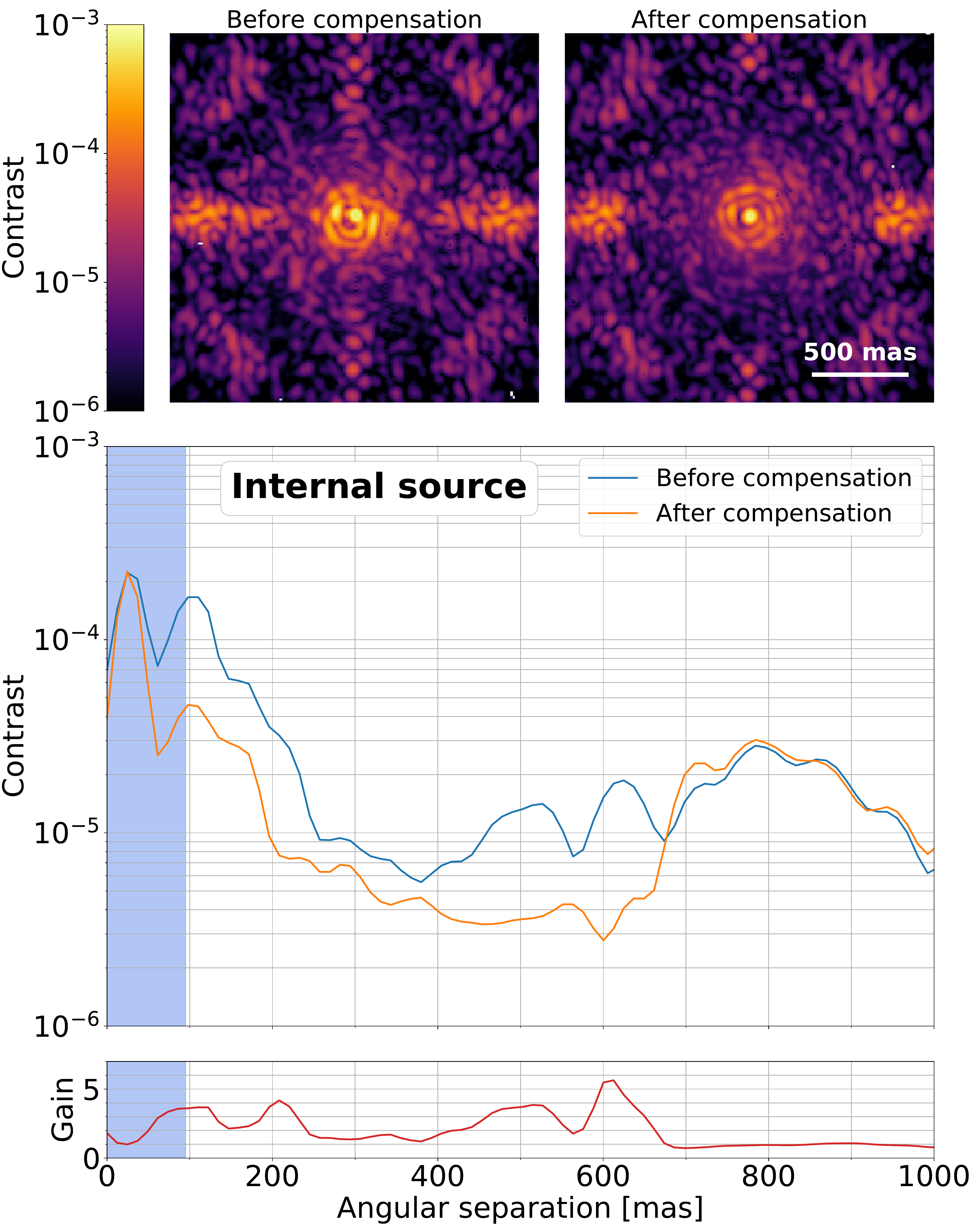}
    \includegraphics[width=0.49\textwidth]{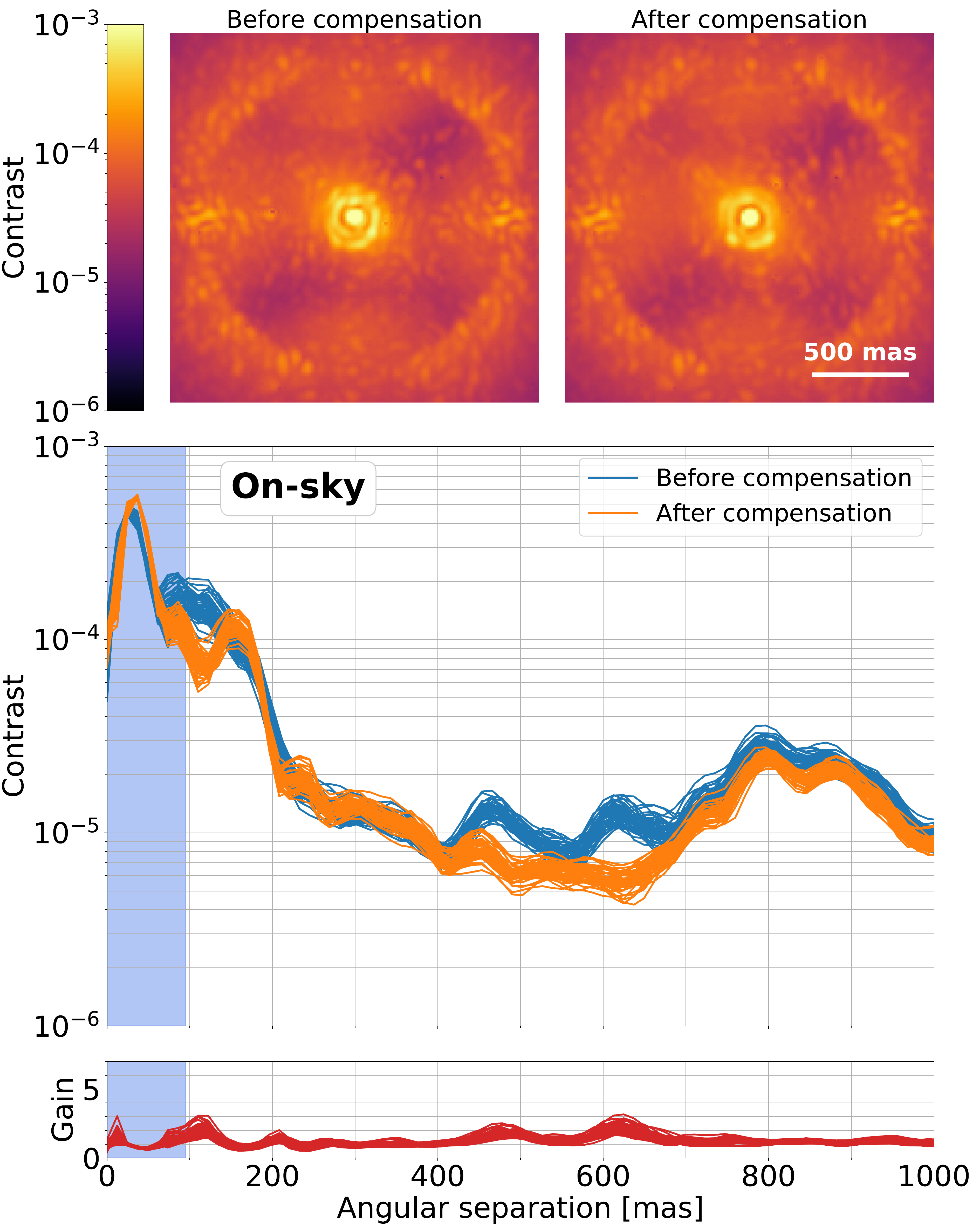}
    \caption{Coronagraphic images and profiles obtained on 2018-04-03 on the internal source (left, test CT03) and on sky (right, test CT04). The top row shows the coronagraphic images before and after NCPA compensation, the main plots show azimuthal standard deviation profiles normalized to the off-axis PSF, and the bottom plots show the contrast gain after NCPA compensation. For on-sky measurements, a profile is shown for each individual coronagraphic image. The blue shadowed region shows the region masked by the focal-plane mask of the APLC.}
    \label{fig:coro_profiles_2}
\end{figure*}

\begin{figure*}
    \centering 
    \includegraphics[width=0.49\textwidth]{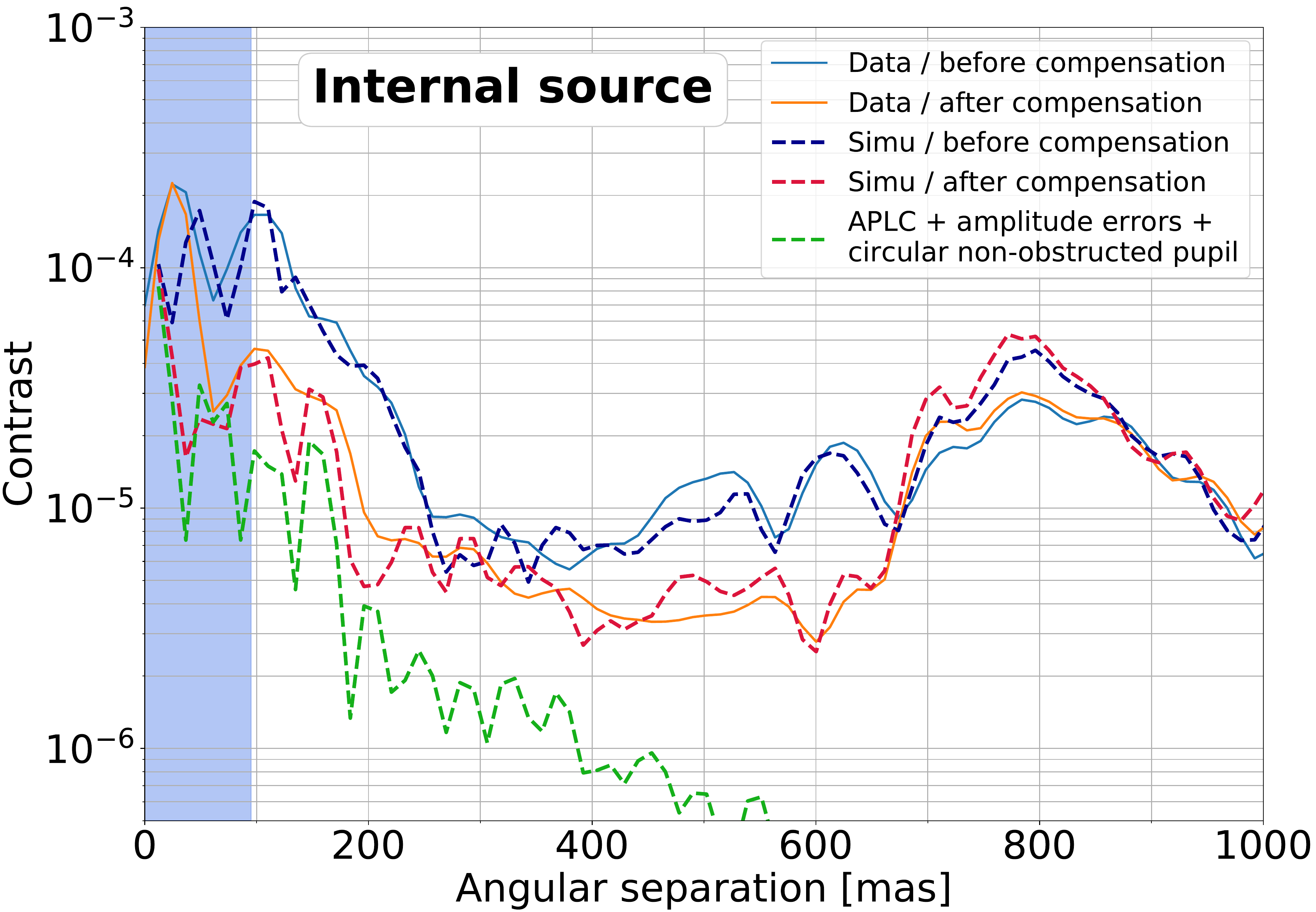}
    \includegraphics[width=0.49\textwidth]{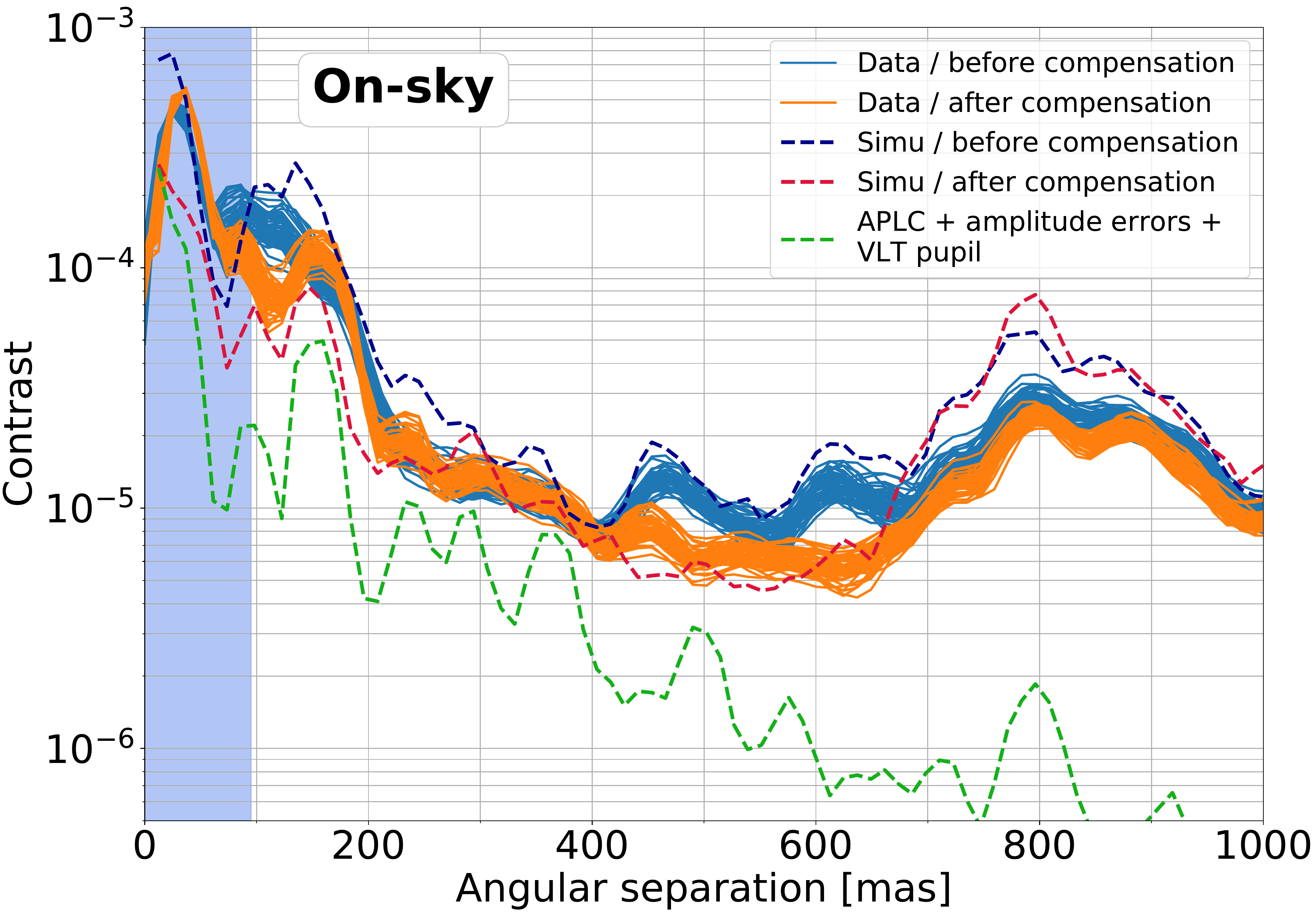}
    \caption{Comparison of the coronagraphic profiles acquired on 2018-04-03 on the internal source (left, test CT03) and on sky (right, test CT04) with coronagraphic image reconstruction based on inputs measured in SPHERE (phase and amplitude aberrations, apodizer transmission, Lyot stop). The theoretical performance of the APLC in presence of amplitude errors is also plotted (dashed green curve). For the on-sky data, the image reconstruction includes ExAO residual wavefront errors based on SAXO real-time telemetry (see Appendix~\ref{sec:phase_screen_reconstruction}). The blue shadowed region shows the region masked by the focal-plane mask of the APLC.}
    \label{fig:coro_profiles_simu_2}
\end{figure*}

This section presents the results on 2018-04-03 for the NCPA compensation loop (Fig~\ref{fig:ncpa_loop_2}), the impact on coronagraphic images (Fig.~\ref{fig:coro_profiles_2}), and the corresponding simulations (Fig.~\ref{fig:coro_profiles_simu_2}). The results on that date are almost identical to the ones that are presented for 2018-04-01 in the main text.

\section{ExAO residual phase-screen reconstruction}
\label{sec:phase_screen_reconstruction}

SAXO enables real-time telemetry to be recorded from the SPARTA real-time computer \citep{Suarez2012}. This recording is however not systematic because of the large amount of generated data (typically $\sim$4~GB/min). The main recorded information is the individual slopes for each of the 1240 sub-apertures of the high-order Shack-Hartmann wavefront sensor. In this section we outline how we used these slopes in combination with the SAXO matrices to reconstruct residual phase screens that are representative of the observing conditions encountered during our tests on sky.

All the HOWFS residual slopes are directly read from the recorded telemetry and stored in a matrix $S$ of dimension $2N_{subap} \times N_{ts}$, where $N_{ts}$ and $N_{subap}$ denote the number of saved time steps and the number of HOWFS sub-apertures. The factor two comes from the storage of the individual $x$ and $y$ slopes. For the reconstruction, we use the following matrices, which are all saved automatically in parallel with the HOWFS slopes when recording real-time telemetry:

\begin{itemize}
    \item $IFM_{\mathrm{HODM}}$: the HODM influence matrix, calibrated from Zygo measurements in the laboratory during the SAXO integration. This matrix is used to go from the HODM voltages to optical path difference in nanometers. It has a dimension of $N_{act} \times 240^2$, where 240 represents the diameter of the pupil in pixels in all SAXO calibrations;
    \item $S2M$: the slope-to-mode matrix that is used to go from orthogonal slope-subspace to KL-mode-space \citep{Petit2008b}. Its dimension is $2N_{subap} \times N_{mode}$.
    \item $M2V$: the mode-to-voltage matrix that is used to go from KL-mode-space to HODM voltage-space. It has a dimension of $N_{mode} \times N_{act}$;
    \item $IM_{\mathrm{ITTM}}$: the image tip-tilt mirror (ITTM) interaction matrix that is used to go from the ITTM voltage-space to the parallel slope-subspace. Its dimension is $2 \times 2N_{subap}$;
    \item $p_{\parallel}$: the parallel projection matrix that is used to go from the HOWFS slope-space to the parallel slope-subspace \citep{Petit2008b}. Its dimension is $2N_{subap} \times 2$;
    \item $p_{\bot}$: the orthogonal projection matrix that is used to go from the HOWFS slope-space to the orthogonal slope-subspace. The orthogonal subspace is defined as the slopes subspace complementary to the parallel slope subspace \citep{Petit2008b}. This matrix has a dimension of $2N_{subap} \times 2N_{subap}$;
\end{itemize}

\noindent with $N_{act}$ being the number of HODM actuators and $N_{mode}$ the number of KL-modes controlled by SAXO. In these matrices, the voltages are normalized to $\pm$1, which means they represent the actuators stroke. 

The SPHERE control architecture includes a specific separation strategy between ITTM and HODM control, first to ensure strict decoupling between the two control loops, and second to allow  dedicated control laws to be used for each control loop. In that framework, the HOWFS slopes space is split into a parallel subspace, corresponding to the slopes subspace addressed by the ITTM actuation. This is a two-dimensional subspace. Its counterpart is $2N_{subap} \times 2$ slopes subspace, and is therefore referred to as orthogonal subspace. The ITTM is controlled from the parallel subspace, while the HODM is controlled from the orthogonal subspace. Consequently, the slopes are first decomposed into their parallel and orthogonal values $S_{\parallel}$ and $S_{\bot}$  using the appropriate projection matrices:
\begin{eqnarray}
    S_{\parallel} & = & p_{\parallel} \cdot S  \\
   , S_{\bot} & = & p_{\bot} \cdot S
.\end{eqnarray}

The tip-tilt and higher orders are then reconstructed separately.

\subsection{Reconstruction of the tip and tilt}

For the tip-tilt reconstruction, some computation is required to obtain the control matrix of the ITTM that is used to go from the HOWFS parallel slope subspace to the ITTM voltage space. This is achieved with
\begin{equation}
IFM_{\mathrm{ITTM}} = \left( p_{\parallel} \cdot IM_{\mathrm{ITTM}} \right)^{-1}\,,
\end{equation}

\noindent where $(\,)^{-1}$ denotes the inverse of the matrix. The ITTM voltages $V_{\mathrm{ITTM}}$ are then reached with
\begin{equation}
    V_{\mathrm{ITTM}} = IFM_{\mathrm{ITTM}} \cdot S_{\parallel}\,,
\end{equation}

and then the respective contribution in tip and tilt to the residual OPD are given by
\begin{gather}
    OPD_{\mathrm{tip}} = Z_1\,D_{tel}\,\tan \left( \gamma V_{\mathrm{ITTM}}
    \begin{bmatrix}
    a_0 & a_2 & a_4 & \ldots
    \end{bmatrix}
    \right),
\end{gather}
\begin{gather}
    OPD_{\mathrm{tilt}} = Z_2\,D_{tel}\,\tan \left( \gamma V_{\mathrm{ITTM}}
    \begin{bmatrix}
    a_1 & a_3 & a_5 & \ldots
    \end{bmatrix}
    \right)\,,
\end{gather}

\indent where $Z_1$ and $Z_2$ represent the 2D Zernike polynomials corresponding to the tip and tilt that are normalized to 1~nm~rms and computed over a  pupil of 240~pixels in size, $D_{tel}$ is the telescope diameter (8~m), and $\gamma = 2.6\as/\mathrm{V} = 1.26 \cdot 10^{-5}\,\mathrm{rad/V}$ is the conversion factor between IM voltages and angular motion on sky. The tip and tilt values are interleaved in the $V_{\mathrm{ITTM}}$ matrix, which is why only the even-index values are used for the tip and the odd-index values are used for the tilt. At the end of this procedure, $OPD_{\mathrm{tip}}$ and $OPD_{\mathrm{tilt}}$ are matrices of size $240^2 \times N_{ts}$. An example of the tip and tilt reconstruction is provided in Fig.~\ref{fig:reconstructed_residual_turbulence}.

\subsection{Reconstruction of the higher orders}

For the reconstruction of the higher orders, we can directly use the available matrices to reconstruct the OPD maps:
\begin{eqnarray}
    M_{\bot}          & = & S2M \cdot S_{\bot} \\
    V_{\mathrm{HO}}   & = & M2V \cdot M_{\bot} \\
    OPD_{\mathrm{HO}} & = & IMF_{\mathrm{HODM}} \cdot V_{\mathrm{HO}} \,.
\end{eqnarray}

At the end of this procedure, $OPD_{\mathrm{HO}}$ is a matrix of $240^2 \times N_{ts}$ in size. An example of the  reconstruction higher orders is provided in Fig.~\ref{fig:reconstructed_residual_turbulence}.

\subsection{Final reconstruction}

\begin{figure*}
    \centering 
    \includegraphics[width=1\textwidth]{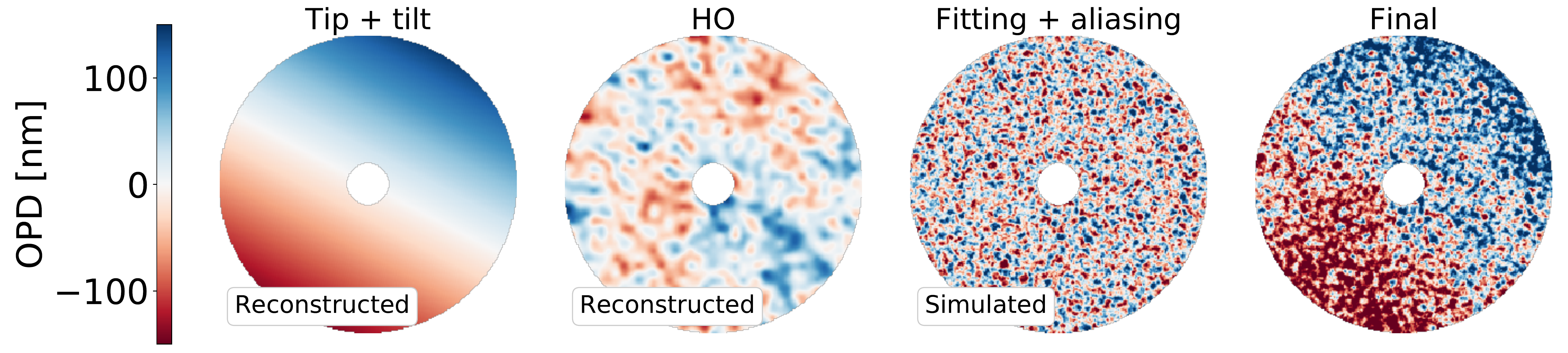}
    \caption{Illustration of the reconstruction of ExAO residual OPD maps at one of the time steps. The reconstruction is decomposed into the reconstruction of the tip and tilt, the reconstruction of the higher orders, and the simulation of the aliasing and fitting error contributions using a Fourier code. The final reconstructed OPD map is the sum of the three previous terms.}
    \label{fig:reconstructed_residual_turbulence}
\end{figure*}

The reconstruction of the OPD maps representing the residuals from SAXO is finally achieved by summing the tip-tilt and higher-order contributions:
\begin{equation}
    OPD_{\mathrm{SAXO}} = OPD_{\mathrm{tip}} + OPD_{\mathrm{tilt}} + OPD_{\mathrm{HO}}\,,
\end{equation}

\noindent which is a matrix of dimension $240^2 \times N_{ts}$. 

However, this does not represent a completely usable residual OPD map because the 40$\times$40 sub-apertures Shack-Hartmann wavefront sensor cannot sense aberrations beyond 20~\cpup. The AO fitting and aliasing errors are therefore not included in the reconstruction. To circumvent these missing contributions, we used a Fourier simulation code based on the modeled power-spectral densities (PSDs) of the AO residuals that are tuned for the SAXO AO system. This code was originally used during the design phase of SAXO \citep{Fusco2006}. This approach presents the advantage of enabling the decomposition of the full AO-filtered PSD into its individual terms, namely the fitting error, the servo-lag error, the aliasing error, the noise error, and the differential refraction.

In the case of our reconstruction, we included the stellar magnitude, zenith distance, and azimuth as input parameters. For the seeing and $C_n^2$ we used the ESO Paranal ambient database to access the data from the MASS\footnote{\url{https://archive.eso.org/wdb/wdb/asm/mass_paranal/form}} \citep{Kornilov2007}, and we estimated that more than 55\% of the turbulence was contained in a ground layer at altitude $z = 0$~km, 35\% in a layer at $z = 4$~km, and the final 10\% in a high-altitude layer at $z = 16$~km. The wind speed and direction of the ground layer were set in the simulation using the values that are also reported in the ambient database.

The Fourier code was then used to generate a PSD that  only includes the fitting error and the aliasing error. For the aliasing, an attenuation factor of 0.5 was used to take into account the fact that SAXO used a spatially filtered Shack-Hartmann wavefront sensor and that the spatial filter was in the \texttt{MEDIUM} position, which only provides a partial attenuation of the aliasing. Using the PSD, we finally generate a series of $N_{ts}$ random OPD maps ($OPD_{\mathrm{Fourier}}$) that are added to the OPD maps reconstructed from the SAXO real-time telemetry:
\begin{equation}
    OPD_{\mathrm{ExAO}} = OPD_{\mathrm{SAXO}} + OPD_{\mathrm{Fourier}}\,.
\end{equation}

\noindent One realization of $OPD_{\mathrm{Fourier}}$ is provided in Fig.~\ref{fig:reconstructed_residual_turbulence}, with the final OPD map $OPD_{\mathrm{ExAO}}$ resulting from the sum of the tip and tilt, the high orders, and the Fourier-simulated aliasing and fitting errors. These reconstructed OPD maps can finally be converted into residual phase screens and injected into numerical simulations to generate images with realistic ExAO residuals (see Appendix~\ref{sec:coronagraphic_image_reconstruction}).

\section{Coronagraphic image reconstruction}
\label{sec:coronagraphic_image_reconstruction}

\begin{figure*}
    \centering 
    \includegraphics[width=1\textwidth]{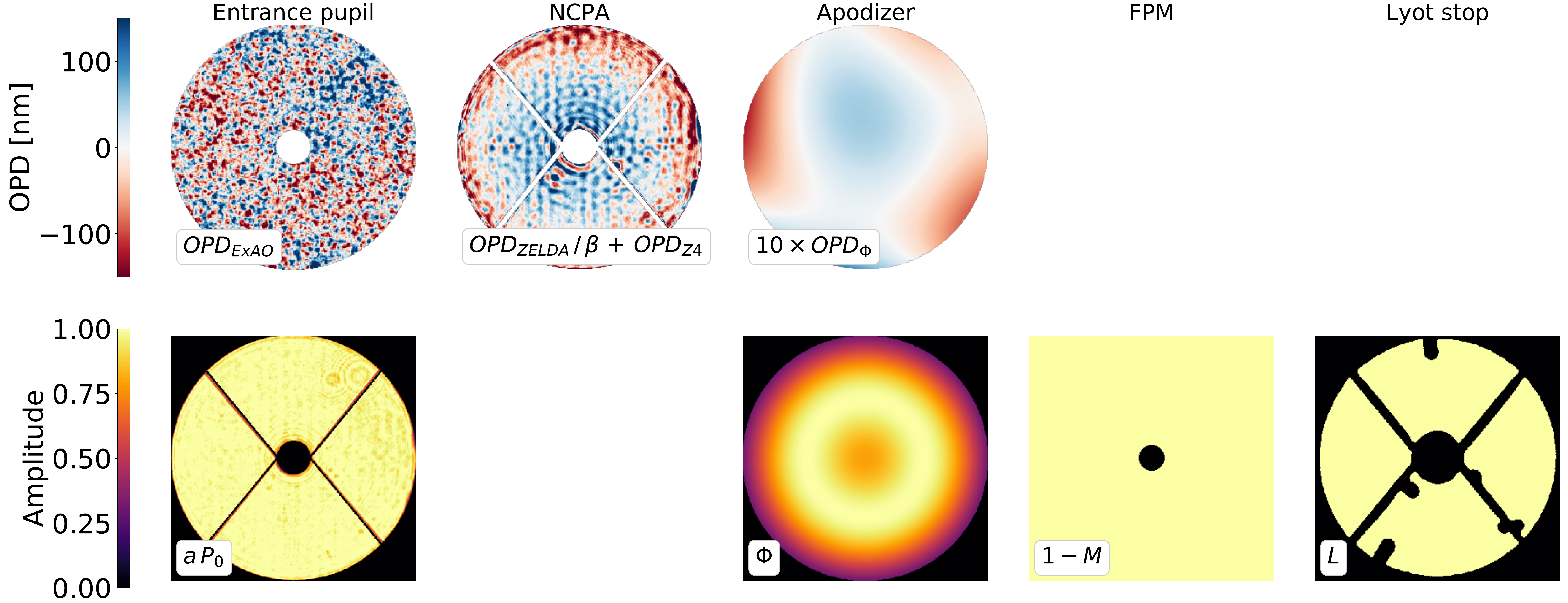}
    \caption{Inputs used in our coronagraphic image reconstruction, illustrated here for the on-sky reconstruction ($\delta_{ExAO} \ne 0$). The OPD contribution of the apodizer ($\delta_{\phi}$) has been multiplied by 10 in this illustration.}
    \label{fig:corono_model_elements}
\end{figure*}

\begin{figure*}
    \centering 
    \includegraphics[width=0.7\textwidth]{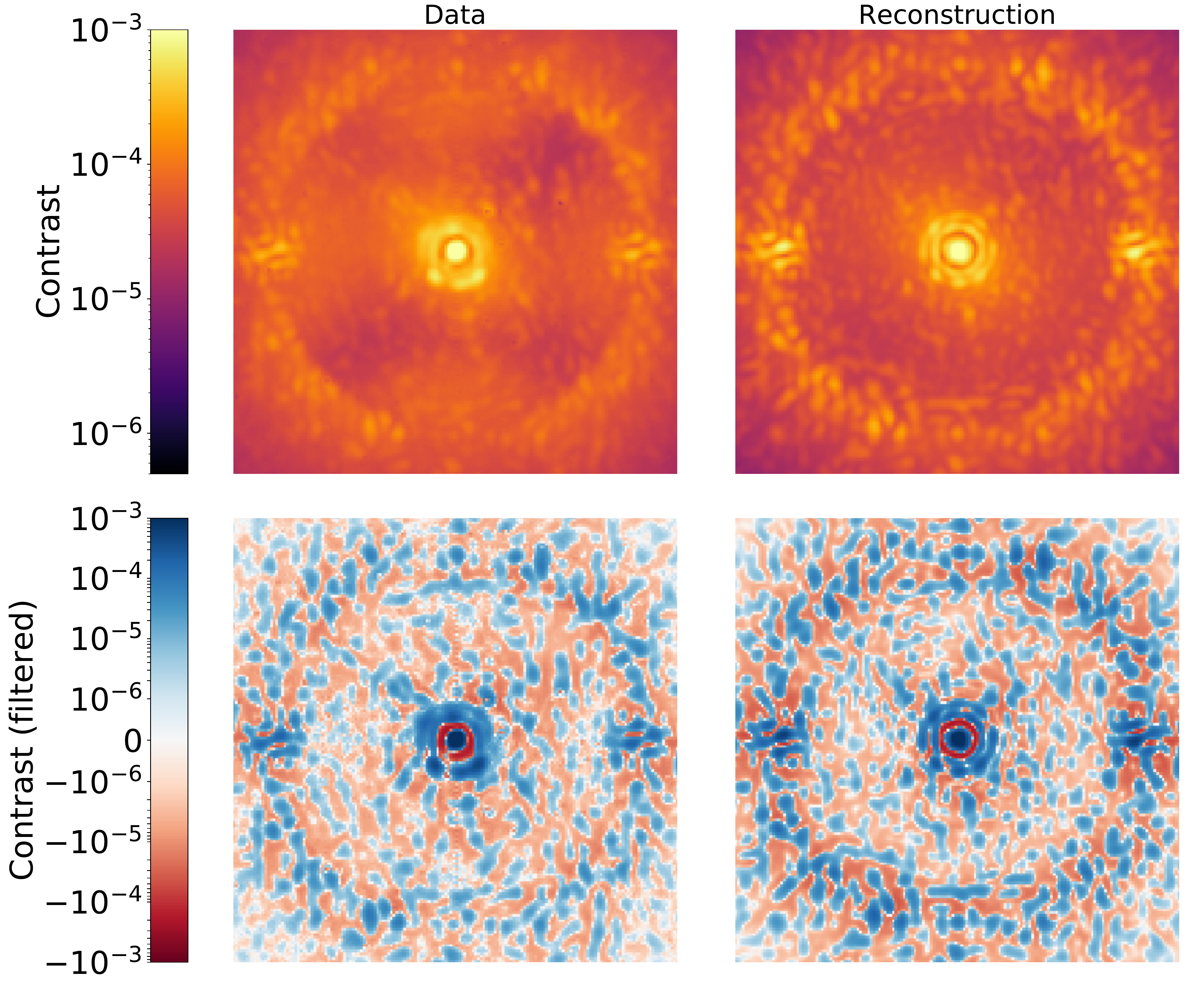}
    \caption{Comparison of the coronagraphic data obtained on sky on 2018-04-03 (left) with the reconstructed coronagraphic image (right). The top row shows the complete data and model, while the bottom row shows a high-pass-filtered version of the images to remove the low-spatial frequencies corresponding to the averaged ExAO residuals.}
    \label{fig:corono_image_reconstruction}
\end{figure*}

We describe the model to reconstruct our broadband coronagraphic images with SPHERE using the APLC parameters and a mix of amplitude and phase errors. While this does not represent a full end-to-end simulation, this model already provides an excellent tool to understand our experimental data on the internal source and on sky, as confirmed by the illustrations shown below.

\subsection{Analytical model}

We briefly reiterate the formalism of the APLC, following the notations of \citet{Aime2002}, \citet{Soummer2003a,Soummer2003b}, and \citet{N'Diaye2016b}. The vectors $\bm{r}$ and $\bm{\xi}$ (of modulus $r$ and $\xi$) denote the two-dimensional position vectors in the pupil and focal planes. Our monochromatic light images are represented at a given $\lambda$ within the spectral bandwidth $\Delta\lambda$ centered at the wavelength $\lambda_0$.

Our simplified model relies on the APLC setup that involves four successive planes
 ($A$, $B$, $C$, $D$) in which $A$ defines the entrance pupil $P$ , which combines the telescope pupil shape $P_0$, the apodization $\Phi$, the amplitude errors $a,$ and phase errors $\varphi$, $B$ sets the location of the focal plane mask, $C$ includes the Lyot stop $L$ to filter out the diffraction due to the coronagraph mask, and $D$ represents the final image plane. The optical layout of the coronagraph is such that the complex amplitudes of the electric field in two successive planes are related by a Fourier transform. We call $\hat{f}$ the Fourier transform of the function $f$. Our coronagraph image reconstruction is performed with the following inputs:
\begin{itemize}
    \item ExAO residual wavefront errors $OPD_{\mathrm{ExAO}}$ reconstructed from SAXO real-time telemetry (see Appendix~\ref{sec:phase_screen_reconstruction}) with a number of maps $N$=41400, corresponding to a 30s integration time sequence with a closed-loop frequency of 1380~Hz;
    \item instrument pupil $P_0$. This is modeled by a clear unobstructed aperture on internal source and a VLT pupil on sky;
    \item amplitude errors map $a$, measured from pupil images in SPHERE;
    \item NCPA OPD map measured with ZELDA $OPD_{\mathrm{ZELDA}}$ and taking into account the loss in sensitivity (see Sect.~\ref{sec:zelda_sensitivity}) due to the ZELDA sensitivity factor $\beta$;
    \item differential defocus term $OPD_{\mathrm{Z4}}$ of -40~nm~rms between the ZELDA mask and the APLC focal-plane mask, estimated from the data (see Sect.~\ref{sec:analysis_limitations});
    \item transmission of the \texttt{APO1} apodizer $\Phi$;
    \item model of the circular hard-edge focal-plane mask of diameter $m$ and transmission $1-M$ with $M(\bm{\xi})=1$ for $\xi < m/2$ and $0$ otherwise;
    \item OPD map of the apodizer $OPD_{\Phi}$, obtained from Zygo measurements during the integration of the instrument. The map used in the simulation is a reconstruction of the measurement with 16 Zernike modes;
    \item transmission of the \texttt{STOP\_ALC2} Lyot stop $L$.
\end{itemize}

These different elements are displayed in Fig.~\ref{fig:corono_model_elements}. As our images were taken in H2 filter on VLT/SPHERE, we use $\lambda_0 = 1.593$~\mic and $\Delta\lambda=0.052$~\mic in our simulations. 

The amplitude of the electric field in the entrance pupil plane A is given by
\begin{equation}
    P(\bm{r}) = P_0(\bm{r})\,\Phi(\bm{r})\,a(\bm{r})\,.\label{eq:P}
\end{equation}
The overall OPD in the entrance pupil plane A is expressed as
\begin{equation}
\begin{split}
    OPD(\bm{r}, k) = & OPD_{\Phi}(\bm{r}) + \frac{1}{\beta}\,OPD_{\mathrm{ZELDA}}(\bm{r})\\ 
    & + OPD_{\mathrm{Z4}}(\bm{r}) + OPD_{\mathrm{ExAO}}(\bm{r}, k)\,, 
    \label{eq:opd}
    \end{split}
\end{equation}
in which $k$ denotes the ExAO residual turbulent phase screen number $k$ out of $N$. The phase error $\varphi$ in the entrance pupil is then deduced from OPD with
\begin{equation}
    \varphi(\bm{r}, \lambda, k) = \frac{2\pi\delta(\bm{r}, k)}{\lambda}\,.
    \label{eq:phase}
\end{equation}

In the absence of magnification, the electric field amplitude in the four successive planes can be expressed as follows
\begin{subequations}
      \renewcommand{\theequation}{\theparentequation\alph{equation}}
\begin{alignat}{2}
& \Psi_A(\bm{r}, \lambda, k)=P(\bm{r})\exp \left (i\varphi(\bm{r}, \lambda, k) \right ) \label{subeq:Psi_A},\\
& \Psi_B(\bm{\xi}, \lambda, k)= \frac{\lambda_0}{\lambda}\widehat{\Psi}_A \left (\frac{\lambda_0}{\lambda}\bm{\xi}, \lambda , k \right ) \left (1- M(\bm{\xi})\right ) \label{subeq:Psi_B},\\
& \Psi_C(\bm{r}, \lambda, k)= \left (\Psi_A(\bm{r}, \lambda, k) - \Psi_A(\bm{r}, \lambda, k)* \frac{\lambda_0}{\lambda}\widehat{M}\left (\frac{\lambda_0}{\lambda}\bm{r}, \lambda \right ) \right )L(\bm{r}) \label{subeq:Psi_C},\\
& \Psi_D(\bm{\xi}, \lambda, k)= \left (\frac{\lambda_0}{\lambda}\right )^2 \left (\widehat{\Psi}_A\left (\frac{\lambda_0}{\lambda}\bm{\xi}, \lambda , k\right ) (1- M(\bm{\xi})) \right )*\widehat{L}\left (\frac{\lambda_0}{\lambda}\bm{\xi}, \lambda \right ), \label{subeq:Psi_D}
\end{alignat}
\end{subequations}
where the asterisk denotes the convolution operator. In the absence of FPM, the electric field amplitude in the final image plane is simply given by
\begin{equation}
\Psi_D^0(\bm{\xi}, \lambda, k)= \left (\frac{\lambda_0}{\lambda}\right )^2 \widehat{\Psi}_A\left (\frac{\lambda_0}{\lambda}\bm{\xi}, \lambda , k\right )*\widehat{L}\left (\frac{\lambda_0}{\lambda}\bm{\xi}, \lambda \right )\,. 
\label{subeq:Psi_0}
\end{equation}

Assuming a flat source spectrum, the direct and coronagraphic image intensity over the band-pass $\Delta\lambda$ can be deduced from Equations (\ref{subeq:Psi_0}) and (\ref{subeq:Psi_D}) as
\begin{subequations}
      \renewcommand{\theequation}{\theparentequation\alph{equation}}
\begin{alignat}{2}
& \mathcal{I}^{0}_{\Delta\lambda}(\bm{\xi})=\frac{1}{N} \frac{1}{\Delta\lambda}\sum_{K} \int_{\Lambda} \left |\Psi^{0}_D(\bm{\xi}, \lambda, k)\right |^2\,d\lambda\\
& \mathcal{I}_{\Delta\lambda}(\bm{\xi})=\frac{1}{N} \frac{1}{\Delta\lambda}\sum_{K} \int_{\Lambda} \left |\Psi_D(\bm{\xi}, \lambda, k)\right |^2\,d\lambda\,,
\label{eq:I_Dbroad}
\end{alignat}
\end{subequations}
where $\Lambda$ defines a set of wavelengths $\lambda$ such that $|\lambda-\lambda_0|< \Delta\lambda/2$ and $K$ defines the set of $N$ ExAO residual phase screens with k ranging from 0 to $N$-1. Finally, our coronagraphic image is normalized to the intensity peak of the direct image $\mathcal{I}^{0}_{\Delta\lambda}$. 

\subsection{Image reconstruction}

We perform our coronagraphic image simulations with the new Python-based object-oriented toolkit called \texttt{Coronagraphs} that enables one to model and optimize Lyot-style coronagraphs and Shaped-pupil devices \citep{N'Diaye2019}. The code will soon be made public and freely available under the MIT license\footnote{\url{https://github.com/astromam/Coronagraphs}}.

For the reconstruction on the internal source, we do not include the ExAO residual atmospheric phase screens and we only simulate a single coronagraphic image. On the contrary, for the on-sky image reconstruction we simulate coronagraphic images with the same number as the ExAO residual phase screens and all the images are averaged in intensity to simulate a long exposure. An example of such a reconstruction for the data acquired on 2018-03-03 is provided in Fig.~\ref{fig:corono_image_reconstruction}. The reconstruction is based on a ZELDA OPD map and pupil amplitude image, which were acquired a few minutes before the coronagraphic data (the ZT04 and CT04 tests are summarized in Tables~\ref{tab:ncpa_loop_log} and \ref{tab:coro_log}), and on the SAXO telemetry that was acquired in parallel with the coronagraphic data. The figure shows both the complete data and reconstruction as well as their high-pass-filtered versions where the low-spatial frequency ExAO residuals have been removed with a simple median filter.

The complete data and reconstruction show strong similarities in the quasi-static speckles structures. However, the data are dominated by aliasing effects in the ExAO-correction region. While the coronagraphic image reconstruction does include an aliasing term that has been added to the reconstruction of the ExAO residual phase screens (see Sect.~\ref{sec:phase_screen_reconstruction}), its modeling does not fully agree with reality. We tried different values for the aliasing attenuation but could never reproduce the same level as in the data. Further work is needed to understand this discrepancy and model the aliasing effect properly in the reconstruction.

To bypass this limitation, we also show a spatially filtered version of the data and the reconstruction in Fig.~\ref{fig:corono_image_reconstruction}. Many structures originating from the quasi-static aberrations are well matched between the data and the model, in particular at the level and beyond the ExAO cutoff. Inside the ExAO-corrected region, there is good correlation of the main speckles but in general the agreement is worse than outside of this region. This means that some aberrations are not properly taken into account in our model. In this very preliminary study we have not yet explored the origin of these discrepancies in full detail, but we can already mention two possible contributors that are not taken into account. The first one is the post-coronagraphic aberrations. Although their contribution is expected to have little effect on coronagraphic images \citep[e.g.,][]{Cavarroc2006}, they cannot be disregarded completely. Unfortunately, ZELDA is only sensitive to the aberrations upstream of the coronagraph, so we currently have no estimation of the downstream aberrations. The second is Fresnel propagation effects inside the instrument, which are going to convert some amplitude aberrations into phase aberrations and vice versa. Our current model based on the classical four-plane APLC setup does not yet take such effects into account. In any case, modeling the Fresnel propagation effects in SPHERE would be extremely challenging without detailed knowledge of the phase and amplitude aberrations that are introduced by each optical component.

Despite these small differences, the agreement is statistically excellent, as demonstrated in the contrast curves that are presented in Figs.~\ref{fig:coro_profiles_simu_1} and \ref{fig:coro_profiles_simu_2}. This is very promising for future work that could use prior knowledge on the instrument to remove part of the ExAO-related structures (e.g., the wind-driven halo, \citealt{Cantalloube2018}) or quasi-static speckles to improve data analysis and interpretation of resolved structures like circumstellar disks.

\end{document}